\documentclass[12pt,preprint]{aastex}

\input{epsf}

\newcommand{\kms}{{\rm km} \, {\rm s}^{-1}}

\newcommand{\gsim}{\mathrel{\hbox{\rlap{\hbox{\lower4pt\hbox{$\sim$}}}\hbox{$>$}}}}
\newcommand{\lsim}{\mathrel{\hbox{\rlap{\hbox{\lower4pt\hbox{$\sim$}}}\hbox{$<$}}}}

\newcommand{\etal}{\rm {et al.} \rm}

\shortauthors{Hoyle \& Vogeley} 
\shorttitle{Voids in the 2dFGRS}

\begin{document}

\title{Voids in the 2dF Galaxy Redshift Survey}

\author{Fiona Hoyle \& Michael S. Vogeley \\ 
Department of Physics, Drexel University, 
3141 Chestnut Street, Philadelphia, PA 19104 \\ 
\email{hoyle@venus.physics.drexel.edu, vogeley@drexel.edu}
}

\begin{abstract}
  We present an analysis of voids in the 2dF Galaxy Redshift Survey
  (2dFGRS). This analysis includes identification of void regions and
  measurement of void statistics. The 2dFGRS is the largest completed
  redshift survey to date, including a total of 245,591 galaxies
  covering 1500 deg$^2$ to a median depth of z$_{\rm med}\sim$ 0.11.
  We use the {\tt voidfinder} algorithm to identify a total of 289
  voids in the 2dFGRS with radius larger than 10$h^{-1}$Mpc. These
  voids have an average effective radius of $14.89\pm 2.65 h^{-1}$Mpc
  in the North Galactic Pole region (NGP) and $15.61\pm 2.84
  h^{-1}$Mpc in the South Galactic Pole region (SGP). These voids are
  extremely underdense, with average density contrast of $\delta \rho
  / \rho= -0.94\pm0.02$. The centers of voids are even emptier,
  because the few galaxies within the voids typically lie close to the
  edges.  The total volume of the universe filled by these void
  regions is approximately 40\%. These results are very similar to
  results found from our analysis of the PSCz survey and the Updated
  Zwicky Catalog; here we detect almost a factor of 10 more voids.  We
  measure the Void Probability Function (VPF) of the 2dFGRS for
  volume-limited samples with limiting absolute magnitudes, M$_{\rm
  lim}$-5log($h$), from $-16$ to $-21$ in b$_J$. We measure the
  Underdensity Probability Function (with density contrast threshold
  $\delta \rho/\rho=-0.8$) for samples with limiting absolute
  magnitudes, M$_{\rm lim}$-5log($h$), from $-18$ to $-21$. We find
  that the SGP is more underdense than the NGP for all but the
  brightest sample under consideration.  There is good agreement
  between the VPF's of the Center for Astrophysics Survey and the
  2dFGRS. Comparison of VPF's measured for the 2dFGRS with the
  distribution of simulated dark-matter halos of similar number
  density indicates that voids in the matter distribution in
  $\Lambda$CDM simulations are not empty enough. However,
  semi-analytic models of galaxy formation that include feedback
  effects yield VPF's that show excellent agreement with the data.

\end{abstract}


\section{Introduction}

In any map of the Universe, there appear large regions that are
avoided by galaxies. These regions may not be completely empty but,
compared to the rest of the Universe, are extremely underdense. We
refer to these regions as voids. The giant void in Bo\"otes was
discovered more than twenty years ago (Kirshner et al. 1981) and the
existence of voids was confirmed by subsequent larger surveys at a
variety of wavelengths (Davis et al. 1992; de Lapparent, Geller, \&
Huchra 1986; da Costa et al. 1988; Geller \& Huchra 1989; Maurogordato
et al. 1992; da Costa et al. 1994; see Rood 1998 and references
therein for a discussion of the history of void detection and
interpretation). The size of the largest voids ($D\sim
30-50h^{-1}$Mpc) in these earlier surveys was of the same order as the
characteristic depth of the surveys, which allowed speculation that
even larger such structures might be found.  Deeper redshift surveys,
for example, the Las Campanas Redshift Survey (Shectman et al. 1996),
confirmed the ubiquity of these structures, but did not detect larger
voids. The recently-completed 2dFGRS (Colless et al. 2003) and the
ongoing Sloan Digital Sky Survey (Abazajian et al. 2003) have the
depth, areal coverage, and complete sampling of the galaxy
distribution necessary to more precisely quantify the distribution of
voids.

The observed spatial distribution of voids and the properties of the
few galaxies that lie within them (Rojas et al. 2003a, b) can strongly
constrain models for cosmology and galaxy formation.  Peebles (2001)
discusses an apparent discrepancy between the cold dark matter model
and observations.  In CDM, voids should be filled with dwarf halos
with escape velocities just larger than the $20\kms$ or so needed to
bind photoionized baryons (Dekel \& Silk 1986; Hoffman, Silk, \& Wyse
1992).  However, surveys of dwarf galaxies indicate that they trace
the same structure as that of other galaxies (Bingelli 1989).  Pointed
observations toward voids also fail to find a significant population
of faint galaxies (Lindner et al.  1996; Kuhn, Hopp, \& Els\"asser
1997; Popescu, Hopp, \& Els\"asser 1997). This is consistent with the
widely-observed result that galaxies have common voids, regardless of
Hubble type (e.g., Thuan \etal 1987; Babul \& Postman 1990; Mo,
McGaugh, \& Bothun 1994). The failure of CDM models to accurately
predict the paucity of dwarfs in voids is connected to the
overprediction of small satellites that would be observed in Milky-Way
sized halos (Kauffmann et al. 1993; Klypin et al. 1999; Moore et al.
1999).  The luminosity function of void galaxies indicates that voids
are not filled with a dense population of void galaxies (Hoyle et
al. 2003).  Semi-analytic modeling of galaxy formation, including
feedback effects, predict that galaxies in voids should be somewhat
fainter, bluer, and more disk-like than galaxies in denser
environments (Benson et al. 2003). Testing these predictions requires
identification of large samples of void galaxies, which in turn
requires detection of many large voids.

To test cosmological models, two complementary approaches may be used
to describe the distribution of voids: detection of voids, i.e.,
treating them as individual structures and studying their properties,
and characterization of large-scale structure using a variety of void
statistics.  A number of techniques have been developed for detecting
voids, including Kauffmann \& Fairall (1991), Kauffmann \& Melott
(1992), Ryden (1995), Ryden \& Melott (1996), El-Ad \& Piran (1997,
EP97), and Aikio \& M\"{a}h\"{o}nen (1998). These algorithms have been
applied to a number of galaxy surveys (Slezak, de Lapparent \& Bijaoui
1993; Pellegrini, da Costa \& de Carvalho 1989; El-Ad, Piran \& da
Costa 1996; El-Ad, Piran and da Costa 1997; M\"{u}ller et al. 2000;
Plionis \& Basilakos 2002; Hoyle \& Vogeley 2002, HV02 hereafter).
The numbers of voids in each sample varies, depending on the
definition of a void, but the void samples in these studies each
contained fewer than 10$^2$ voids. In this paper we describe detection
of more than 200 voids in the 2dFGRS.

The distribution of voids may also be characterized using statistics
such as the Void Probability Function (VPF, White 1979) and the
Underdensity Probability Function (UPF, Vogeley et al. 1989),
which depend on the hierarchy of n-point correlation functions. The
VPF is simply the probability that a randomly selected volume contains
no galaxies. The UPF measures the frequency of regions with density
contrast $\delta \rho /\rho$ below a threshold. These statistics
reveal statistical information about the void population but do not
give details on specific voids.  Halo-occupation models of galaxy
formation (e.g., Berlind \& Weinberg 2002) show that statistics of
voids can strongly constrain models for galaxy biasing models.
Previous work on the VPF includes examination of the sky-projected
galaxy distribution (Sharp 1981; Bouchet \& Lachi\`eze-Rey 1986),
galaxy redshift surveys (Hamilton, Saslaw, \& Thuan 1985; Maurogordato
\& Lachi\`eze-Rey 1987, 1991;  Fry \etal 1989; Mo \& B\"orner 1990;
Vogeley, Geller, \& Huchra 1991; Lachi\`eze-Rey, da Costa, \&
Maurogordato 1992; Bouchet \etal 1993; Vogeley et al.
1994), and the distribution of clusters of galaxies (Huchra \etal
1990; Jing 1990; Cappi, Maurogordato, \& Lachi\`eze-Rey 1991). 

Void statistics measured for the CfA2 survey and simulations of CDM
models indicate that the models can reproduce voids of bright
galaxies, but fail to match the statistics of fainter galaxies
(Vogeley \etal 1994; Cen \& Ostriker 1998; Benson et al. 2003).  These
statistics show that voids in the simulations can actually appear too
empty.  This result strongly conflicts with the visual impression that
observed voids have very smooth edges, while structures in mock
redshift surveys from simulations are more diffuse (e.g., Diaferio
\etal 1999).  The large difference between voids in the dark matter
and galaxy distributions (Ostriker et al. 2003) strongly suggests that
the inconsistency lies in the details of treating galaxy formation in
voids.  Results of high-resolution hydrodynamic/N-body simulations
show strong density dependence of the efficiency of galaxy formation
(Blanton et al. 1999; Ostriker et al. 2003).  Precise measurement of
the frequency of voids in samples of varying luminosity and type will
yield strong tests of the details of galaxy formation models.

In this paper, we apply the {\tt voidfinder} analysis of HV02 and the
VPF and UPF to the largest completed galaxy redshift survey, the 2
degree Field Galaxy Redshift Survey. This survey contains $\sim
250,000$ galaxies over 1,500 deg$^2$. We describe the data in more
detail in section \ref{sec:survey}. In section \ref{sec:algo}, we give
an overview of the {\tt voidfinder} algorithm, present the results in
section \ref{sec:res}.  In section \ref{sec:vpf} we turn our attention
to the VPF and UPF, give a brief outline of these algorithms and
present our results on the VPF and UPF
statistics . In section \ref{sec:conc} we summarize our results
and present our conclusions.

\section{The 2dFGRS}
\label{sec:survey}

The 2dFGRS is an optical spectroscopic survey of objects brighter than
b$_{\rm J} \sim$19.30 selected from the APM Galaxy Survey (Maddox et
al. 1990a, b). The effective median magnitude limit of the survey is
19.30, lowered from the original magnitude limit of b$_{\rm J}=$19.45
because the photometry of the input catalog and the dust extinction maps
have been revised so there are small variations in the magnitude limit
as a function of position over the sky (Colless et al. 2003).

The 2dFGRS survey is divided into two main regions, with additional
random fields observed to improve the angular coverage for statistics
such as the power spectrum. For this analysis, we do not use the
random fields as we can only find voids in regions that have a wide
angular range. The two areas of interest are the South Galactic Pole
(SGP) region, which covers the region $ 325^{\circ} < \alpha <
52.5^{\circ}$, $-37.5^{\circ} < \delta < -22.5^{\circ}$ and a region
which lies toward (but not terribly close to) the North Galactic Pole
(NGP), $ 147.5^{\circ} < \alpha < 222.5^{\circ}$, $-7.5^{\circ} <
\delta < 2.5^{\circ}$. The data that we analyze are from the public
release that was distributed to the community in July 2003 (Colless et
al. 2003 and references therein). 245,591 galaxy redshifts were
contained in the data release, making this the largest completed
redshift survey to date.

The completeness of the survey varies with position on the sky
because of unobserved fields (mostly around the survey edges),
observed objects with poor spectra, and objects that could not be observed
due to either fiber collision constraints or broken fibers. To match
the angular selection function in the construction of the random
catalogs, required in our void finding algorithm, we use the software
developed by the 2dFGRS team and distributed as part of the data
release \footnote{for the data release products and catalogs see {\tt
www.mso.anu.edu.au/2dFGRS}}. For any given coordinates ($\alpha,
\delta$), the expected probability of a galaxy being contained in the
2dFGRS survey region is returned.

We construct volume-limiteds sample from the 2dFGRS to obtain a
uniform radial selection function, thus the only variation in the
space density of galaxies with radial distance is due to clustering.
This means the chance of finding a void at any given distance in such
a sample should depend only on the local clustering rather than on the
changing selection function.
We adopt the global $k$-correction + evolution correction found for
b$_{\rm J}$ selected galaxies in the ESO Slice Project (Zucca et al.
1997) and adopted by the 2dFGRS team (Norberg et al. 2001) and use
\begin{equation}
k + e = \frac{0.03 z}{(0.01 + z^4)}.
\end{equation} 
For voidfinding, we select a volume limit of $z_{\rm max}$=0.138. This
value maintains the maximum number of galaxies in the sample: 27,573
galaxies in the NGP region and 33,581 galaxies in the SGP region.  We
adopt a $\Omega_{\rm m}=0.3, \Omega_{\Lambda}=0.7$ cosmology when
converting redshift into comoving distances. For this cosmology, a
redshift limit of $z_{\rm max}$=0.138 corresponds to a comoving
distance of 398$h^{-1}$Mpc.  For measuring void statistics, we examine
sub-samples with a range of absolute-magnitude limits.

\section{The Void Finding Algorithm}
\label{sec:algo}

\subsection{Outline}

The void finding algorithm we adopt is described in full detail in
HV02, including justification of each parameter value. The method is
similar to that of EP97. The steps of the void
finding algorithm are as follows:
\begin{itemize} 
\item Classification of galaxies as wall or void galaxies
\item Detecting empty cells in the distribution of wall galaxies
\item Growth of the maximal empty spheres
\item Classification of the unique voids
\item Enhancement of the void volume
\end{itemize}

Following HV03, we calculate the mean distance, {\it d}, to the
$3^{\rm rd}$ nearest galaxy. Any galaxy that has less than three
neighbors in a sphere of radius {\it l$_3$} = 5.6$h^{-1}$Mpc is
considered a void galaxy. The remaining galaxies are labeled as wall
galaxies. We note that the value of $l_3$ is similar to the value used
in Rojas et al. (2003a) to identify void galaxies in the SDSS.

The next step is to place the wall galaxies onto a three dimensional
grid. Each empty grid cell is considered to be part of a possible
void. Our method finds the maximal sphere (hole) that can be drawn in
the void starting from the empty cell. We insist that the holes lie
completely in the survey. At this stage we keep track of all the holes
with radii larger than the value of the search radius used to classify
void and wall galaxies, $l_{3}$. The fineness of the grid defines the
minimum size void that can be detected. All holes with size larger
than r = $\sqrt{3}$l$_{\rm cell}$, where l$_{\rm cell}$ is the size of
each grid cell, will be detected. We use a cubical mesh that has 128
grid cells on a side, each of which is 4.7h$^{-1}$Mpc in length, thus
we find all holes larger than 8.1$h^{-1}$Mpc.

Finding the holes is a robust process. Deciding which of those holes
are unique voids requires more thought. First we sort the holes by
radius, the largest first. The largest hole found is automatically a
void. We examine whether the second largest hole overlaps the void of
not. If it overlaps by more than 10\% in volume we say this hole is
part of the first void, if not then it forms a new void. We continue
like this for all holes with radii larger than 10$h^{-1}$Mpc. 
However, if a hole overlaps more than one previously detected void 
by more than 10\%, the hole is ignored as it links together two 
larger, previously identified voids. 

We next enhance the volume of each void. Any hole that overlaps the
maximal void sphere by 50\% of the smaller hole's volume is
considered part of the larger void. If the hole overlaps with more
than one void then it is not added to either of the voids as this
again would link two voids together that we wish to keep separate. If the
hole is isolated and less than 10$h^{-1}$Mpc in radius, it cannot be
classified as a separate void as we only call a hole a void if it is 
larger than 10$h^{-1}$Mpc in radius.

Finally we count the number of galaxies that lie within each void to
determine the underdensity of the void. Measured properties of each
void include the coordinates of its center, the maximal sphere radius,
the effective radius, and the average underdensity.

\section{Properties of Voids}
\label{sec:res}

\begin{figure} 
\begin{centering}
\begin{tabular}{c}
{\epsfxsize=14truecm \epsfysize=14truecm \epsfbox[40 120 600 700]{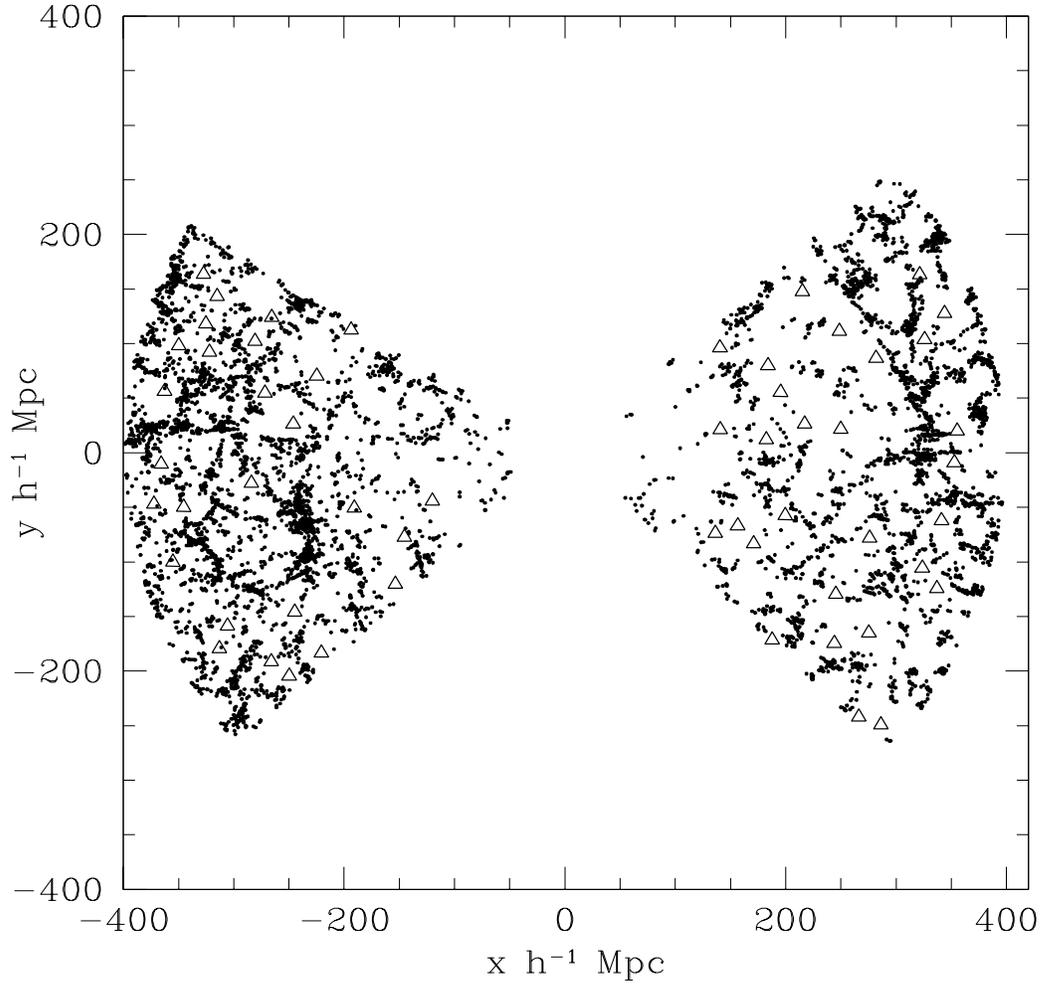}} \\
\end{tabular}
\caption{Voids in the NGP and SGP. Here we show an example of the
distribution of wall galaxies (filled points) and the centers of voids
(open triangles) in thin ($1^{\circ}$) slices of the NGP and SGP.}
\label{fig:voidspic}
\end{centering}
\end{figure}

We apply the {\tt voidfinder} algorithm to volume-limited samples of
the 2dFGRS and find 116 voids in the NGP region and 173 voids in the
SGP region. The properties of all the voids are available upon request
and are summarized in Table \ref{tab:voidprops}. Figure
\ref{fig:voidspic} shows an example of the distribution of wall
galaxies and the centers of voids in thin ($1^{\circ}$) slices of the
NGP and SGP. The filled circles show the location of the wall
galaxies, the open triangles mark the centers of the voids. The SGP
region covers an area that is approximately 33\% larger than the NGP
but it contains $\sim 50\%$ more voids. In this section we examine
these voids in detail and compare the results for the NGP and SGP
regions.

\subsection{Significance of the Voids}
\label{sec:conf}

\begin{figure} 
\begin{centering}
\begin{tabular}{c}
{\epsfxsize=8truecm \epsfysize=8truecm \epsfbox[40 120 600 700]{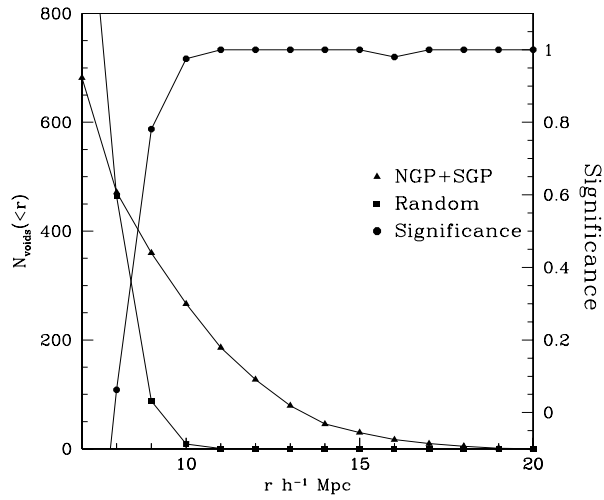}} \\
\end{tabular}
\caption{Statistical significance of voids. Triangles show the cumulative
distribution of voids in the NGP and SGP, extended below a maximal
sphere radius of 10$h^{-1}$Mpc for this purpose only. Squares show
the cumulative distribution of voids in random samples, averaged over 10
realizations. Circles show the void significance at each scale. All voids
larger than 10$h^{-1}$Mpc are significant at the 90\% level. On smaller scales
the significance drops rapidly.}
\label{fig:voidsig}
\end{centering}
\end{figure}

To assess the statistical significance of the detected voids, we
generate 10 Poisson samples of the 2dFGRS that have the same angular
selection function, distant limits, and number density as the 2dFGRS
but contain no clustering.  The statistical significance of a void may
be estimated from (El-Ad, Piran \& da Costa 1997)
\begin{equation}
p(r) = 1 - \frac{N_{\rm Poisson}(r)}{N_{\rm Survey}(r)}
\end{equation}
where $N_{\rm Poisson}(r)$ and $N_{\rm Survey}(r)$ are the numbers of
voids detected by applying {\tt voidfinder} to random samples and galaxy
samples, respectively.
In Figure \ref{fig:voidsig}, we show
the number of voids with radius greater than $r$ that we find in the NGP
and SGP added together (triangles) as compared to the random catalogs
(squares). In the same plot, we present the significance of the voids.
Voids larger than 10$h^{-1}$Mpc are significant at the $>90$\% level,
consistent with results in HV02. The statistical significance of voids 
rapidly approaches 100\% as the void radii increase;
the false positive rate for voids with average radius $r\sim
12h^{-1}$Mpc is well below 1\%.

\begin{table}
\begin{centering}
\begin{tabular}{cccccc}
Region & N$_{\rm voids}$ & r ($h^{-1}$Mpc) & r$_{\rm eff}$ ($h^{-1}$Mpc) & $\delta \rho / \rho$ & Volume \\ \hline
NGP & 116 & 12.09$\pm 1.85 h^{-1}$Mpc & 14.89$\pm 2.65 h^{-1}$Mpc & 
-0.94$\pm0.02$ & 33.4\% \\
SGP & 173 & 12.52$\pm 1.99 h^{-1}$Mpc & 15.61$\pm 2.84 h^{-1}$Mpc & 
-0.93$\pm0.02$ & 41.6\% \\ \hline
\end{tabular}
\caption{The properties of the voids in the NGP and SGP}
\label{tab:voidprops}
\end{centering}
\end{table}

\subsection{Sizes of the Voids}
\label{sec:sizevoid}

\begin{figure} 
\begin{centering}
\begin{tabular}{cc}
{\epsfxsize=8truecm \epsfysize=8truecm \epsfbox[40 120 600 700]{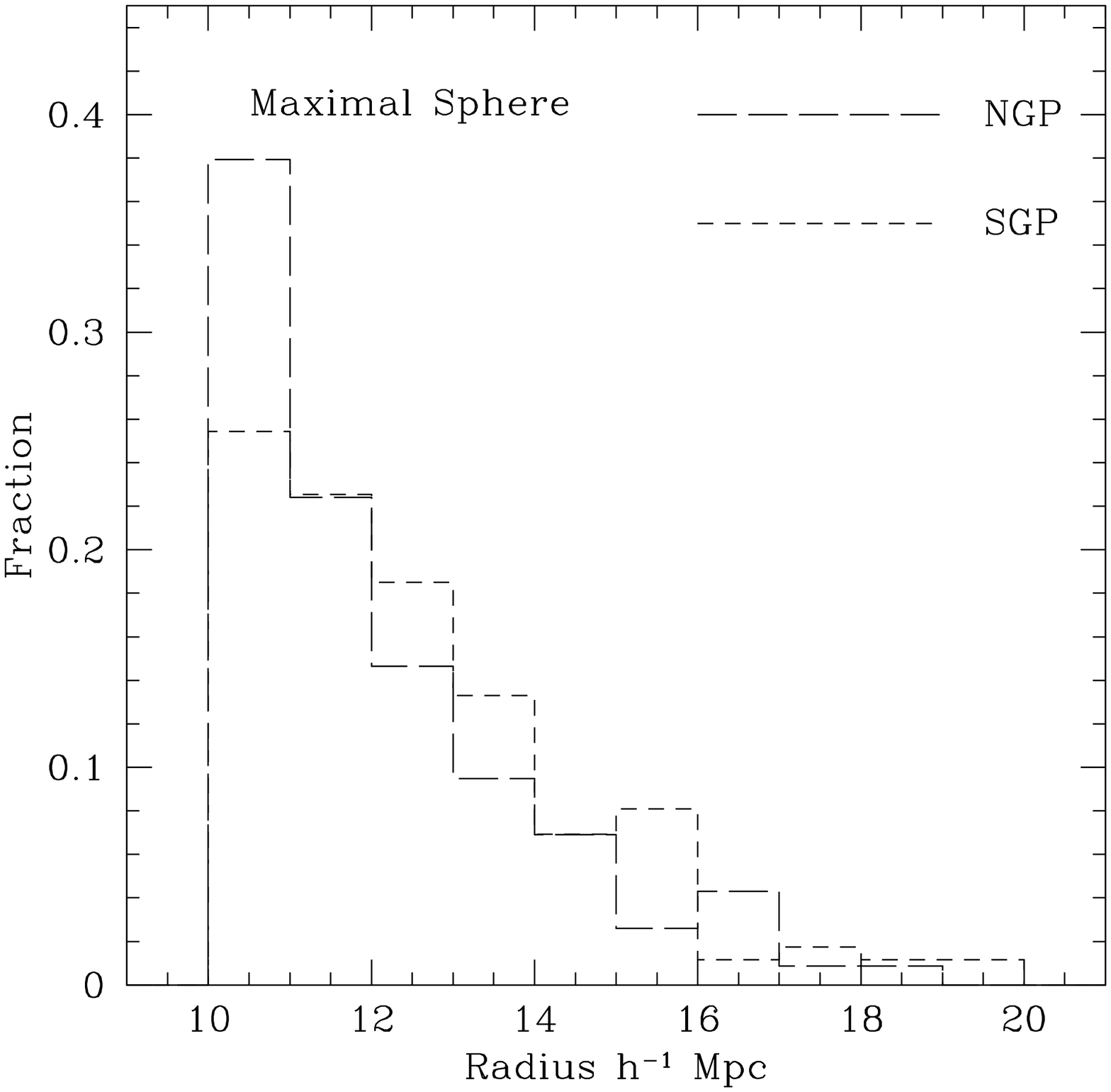}} &
{\epsfxsize=8truecm \epsfysize=8truecm \epsfbox[40 120 600 700]{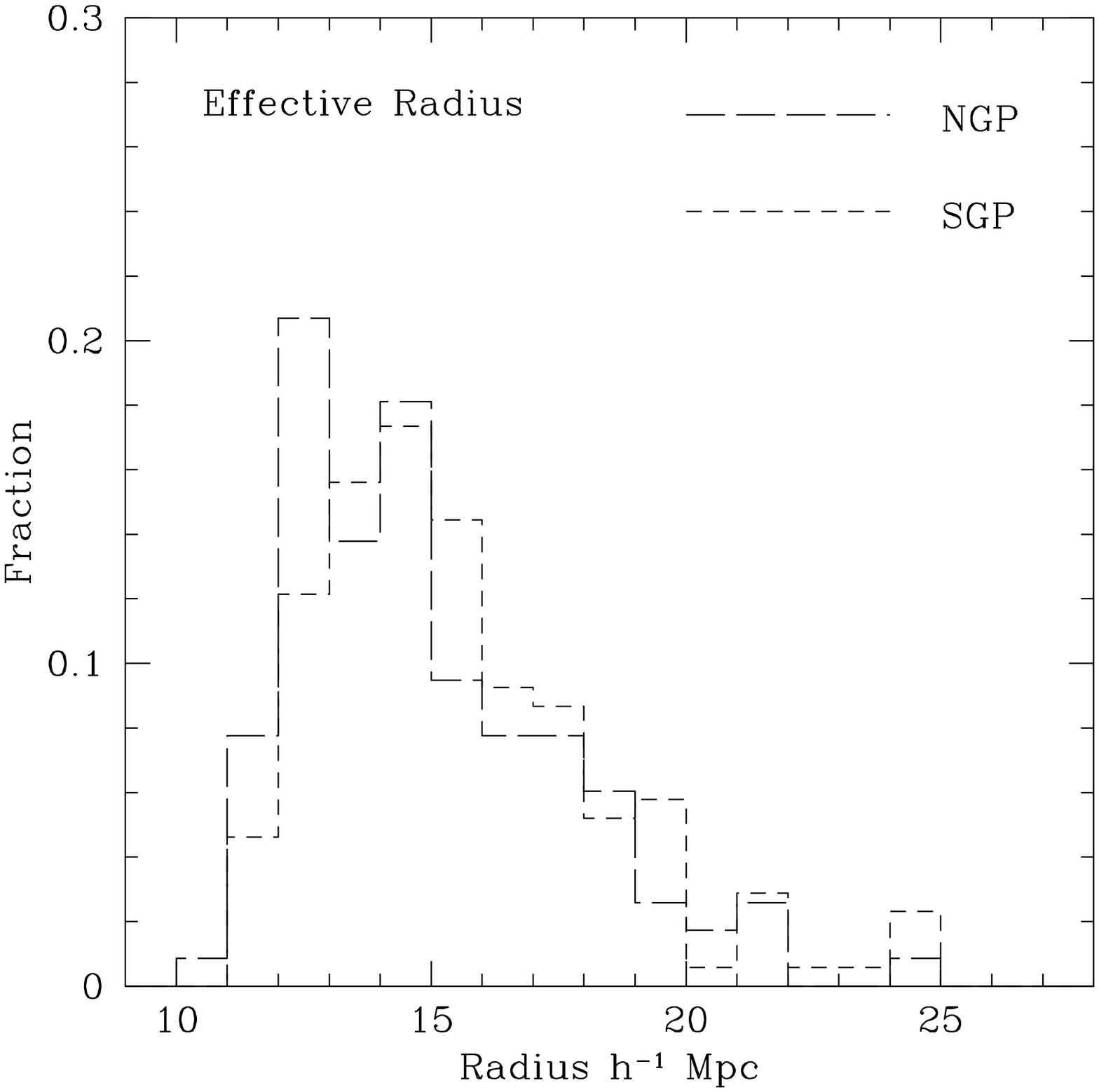}} \\
\end{tabular}
\caption{Distribution of maximal sphere sizes and effective radii
for the voids in the NGP (long dashed line) and the SGP (short dashed
line). The average maximal sphere size is $12.09 \pm 1.85 h^{-1}$Mpc
for the NGP and $12.52 \pm 1.99 h^{-1}$Mpc for the SGP and the average
effective void size is $14.89 \pm 2.65 h^{-1}$Mpc for the NGP and
$15.61 \pm 2.84 h^{-1}$Mpc for the SGP.}
\label{fig:voidsize}
\end{centering}
\end{figure}

The mean maximal sphere size, with minimum size of 10$h^{-1}$Mpc, is
$12.09 \pm 1.85 h^{-1}$Mpc for the NGP and $12.52 \pm 1.99 h^{-1}$Mpc
for the SGP so the two samples of voids have similar average sizes,
although the voids in the SGP are slightly larger. Average values of
effective radius are slightly larger, $14.89 \pm 2.65 h^{-1}$Mpc for
the NGP and $15.61 \pm 2.84 h^{-1}$Mpc for the SGP. Figure
\ref{fig:voidsize} shows histograms of the distribution of void sizes.
The distribution of the maximal spheres are shown in the left and the
effective radii on the right. The NGP distribution peaks at lower
radii, suggesting that the SGP region is slightly emptier.
Statistical comparison of the distributions of void sizes using a
Kolmogorov-Smirnov test reveals that the NGP and SGP void size
distributions are different only at the 6\% level, thus the volume of
these samples are just large enough to overcome variations of the
large-scale structures within them (a.k.a., cosmic variance).

\subsection{Density of Voids}
\label{sec:voidrho}

The galaxies in the NGP and SGP samples are split into
wall/void galaxies as described in section
\ref{sec:algo}. Approximately 10\%
of galaxies lie in low density regions, are classified as void
galaxies, and could reside in the void regions. However, due to the
shapes of the voids, some of the void galaxies do not lie within the
detected voids, but rather lie at the edges near a higher density region.
In the NGP, 1518 of the 2296 void galaxies reside
in a void and in the SGP, 2319 of the 3992 void galaxies are found in
a void, thus 5.5\% and 6.9\% of the galaxies lie in voids. These results
are consistent with our analysis of the UZC and PSCz (HV02) and 
results of El-Ad, Piran \& da Costa (1997). 

The 5-7\% of galaxies that lie within voids inhabit extremely rarified
regions of the universe.  The average density contrast of voids is
$\delta\rho/\rho=-0.94\pm0.02$ and $-0.93\pm0.02$ for the NGP and SGP,
respectively. If we examine the density profiles of voids in detail,
we find that the centers of voids are even more empty, with density
contrast approaching $\delta\rho/\rho=-1$, i.e., zero density of
galaxies.  To consistently compare the density profiles of voids of
varying size, we normalize distances from the void center by the
maximal sphere size for each voids. Thus, we measure density as a
function of $r/R_{max}$.  In other words, a galaxy lying 6$h^{-1}$Mpc
from the center of a 12$h^{-1}$Mpc void is considered to be at the
same `distance' as a galaxy 8$h^{-1}$Mpc from the center of a
16$h^{-1}$Mpc void. The average effective radius for the voids is
$\sim 15 h^{-1}$Mpc.  We calculate the density of the void assuming
this average size and compare the average density of the two regions
to calculate $\delta \rho / \rho$.

\begin{figure} 
\begin{centering}
\begin{tabular}{c}
{\epsfxsize=8truecm \epsfysize=8truecm \epsfbox[40 120 600 700]{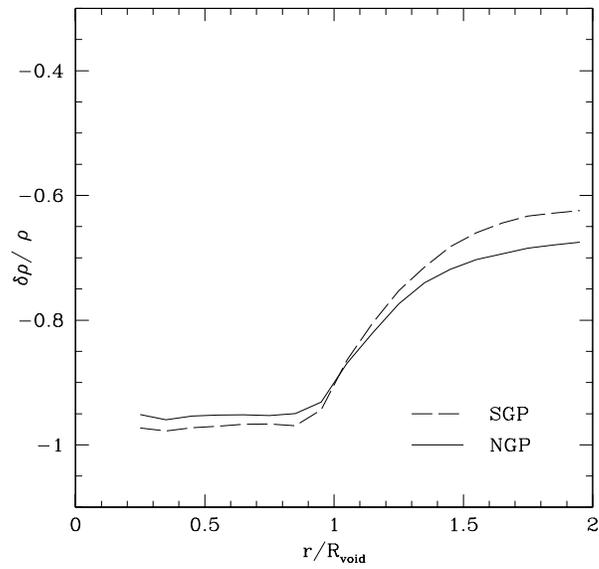}} \\
\end{tabular}
\caption{Radial density profile of voids.  Within the void radius
($r/R_{\rm void}<1$), the voids are very empty, with density contrast
($\delta \rho / \rho < -0.9)$.
Just beyond the
maximal sphere radii the density of the voids rises
rapidly. However, even at twice the void radius, the density remains
lower than average.}
\label{fig:voidprofile}
\end{centering}
\end{figure}

In Figure \ref{fig:voidprofile}, we show the average density contrasts
within voids in the NGP (solid line) and SGP (dashed line).  The
voids are extremely underdense in the centers.  Even at 90\% of the
void radius the density contrast remains below $\delta\rho/\rho=-0.9$. The
density of the void then sharply rises at the maximal sphere radius.
indicative of the fact that we have accurately found the
walls surrounding the voids.
Even at twice the void radius, the voids still trace underdense
regions. The same trend is seen in both the NGP and the SGP. The
profiles of the voids show how well the void finder algorithm works
in detecting empty regions. The 10\% or so of galaxies that are
allowed to reside in void regions tend to lie close to the edges of
the voids. 

Tests of {\tt voidfinder} using cosmological simulations also show how
well this algorithm works. Using semi-analytic models applied to CDM
simulations, Benson et al. (2003) created mock galaxy catalogs with
the same density as the 2dFGRS and SDSS. When applied to these mock
samples, {\tt voidfinder} accurately detected the edges of
voids. Comparison of the density profiles of the voids in the mock
galaxies with the density profiles of dark matter showed that the
sharp transition from void to wall was seen in both galaxies and dark
matter. 

\subsection{Void Size as a Function of Redshift}

The $\Lambda$CDM model with parameters inferred from the WMAP
measurements of the CMB anisotropy (Bennett et al. 2003) is able to
reproduce the large-scale structure of the universe. On small scales
there are some problems, as the models produce too much substructure,
but perhaps this problems can be addressed by including the physics of
galaxy formation.  On the scale of voids, this basic model for
formation of structure via gravitational instability in a universe in
which gravitational potential fluctuations are dominated by dark
matter seems fairly secure. During the evolution of large-scale
structure, voids deepen and expand in comoving size. We might,
therefore, be able to detect this growth if we can examine void sizes
over a sufficient range of redshift.

\begin{figure} 
\begin{centering}
\begin{tabular}{c}
{\epsfxsize=8truecm \epsfysize=8truecm \epsfbox[40 120 600 700]{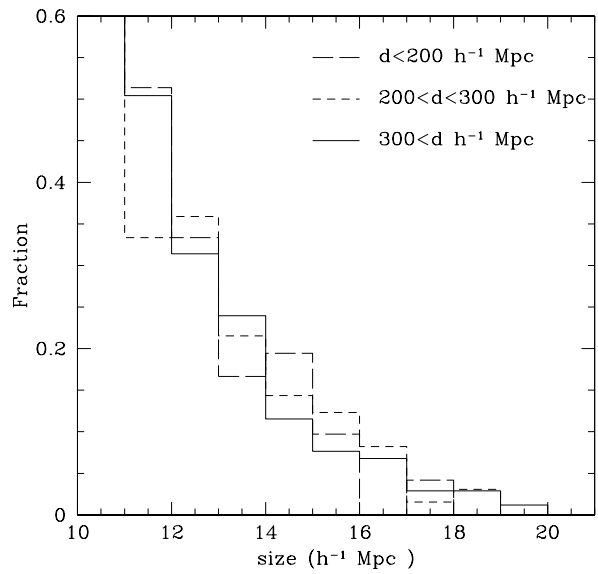}} \\
\end{tabular}
\caption{Test for evolution of void sizes. Lines show the fraction of voids
as a function of the maximal sphere radius split into three groups by
comoving distance: $<200 h^{-1}$Mpc (long-dash), $200-300 h^{-1}$Mpc (short
dash) and $>300 h^{-1}$Mpc (solid). There is little
difference in the histograms, hence no indication of evolution in the
sizes of voids over the range of redshift covered by the 2dFGRS.}
\label{fig:voidredshift}
\end{centering}
\end{figure}

The volume-limited samples we extract from the 2dFGRS extend to
z$_{\rm max}$=0.138, which corresponds to a comoving distance of $\sim
400 h^{-1}$Mpc.  To make a first test for the variation with redshift
of void size, we divide our void sample into three ranges of comoving
distance and compare the histograms of void sizes.  In Figure
\ref{fig:voidredshift}, we combine the voids in the NGP and SGP and
compare the distributions of voids whose centers lie at $ r < 200
h^{-1}$Mpc (long dashed line), $200 < r < 300 h^{-1}$Mpc (short dashed
line) and $300 < r h^{-1}$Mpc (solid line).  If void sizes grow with
time, we would expect the nearby sample to contain larger voids.
However, here we find that the most distant samples contain the largest
voids.  Clearly, the 2dFGRS does not extent deep enough to observe
the growth of voids.  To leading order, structure growth freezes out at a
redshift of $\sim 1/\Omega_{\rm m}$ which is around $z\sim 3$ for a
$\Omega_{\rm m}=0.3$, $\Omega_{\Lambda}=0.7$ cosmology, so little
evolution could be detected in the samples here.

\subsection{Volume of the Universe Contained in Voids}

We estimate the total volume of the 2dFGRS contained in the voids
using random catalogs which have the same angular and radial selection
function as the volume limited catalogs but the particles are
unclustered.  Each random particle is tested as to whether it lies in
a void region or not and the volume contained in voids is the simply
the number of random particles that lie in the voids, divided by the
total number of random particles.  This procedure is essentially a
Monte Carlo integration over the strangely-shaped volumes of all the
voids (recall that the final merged voids are not exactly spherical).
We find that 33.4\% of the NGP volume is contained in a void and
41.6\% of the SGP region is contained in a void. This is consistent
with early findings that the SGP is emptier than the NGP.  Note
carefully that the fraction of space filled by voids depends strongly
on the definition of ``void.'' As shown by Figure
\ref{fig:voidprofile}, we count only the deep interior of voids,
excluding the transition zone near the dense structures.  Thus, these
voids are structures with extremely low mean density,
$\delta\rho/\rho<-0.9$.

\subsection{Comparison with Voids in the UZC and PSCz}

Voids have previously been detected in the Updated Zwicky Catalog and
the PSCz survey using {\tt voidfinder} (HV02) and other authors have
applied different void finding methods to detect voids in other
surveys (see, for example Slezak, de Lapparent \& Bijaoui 1993;
Pellegrini, da Costa \& de Carvalho 1989; El-Ad, Piran \& da Costa
1996, 1997; M\"{u}ller et al. 2000; Plionis \& Basilakos 2001).  It
would be interesting to compare the void catalogs using different
methods, but because of the narrow opening angle of the 2dFGRS, it is
not possible to detect voids in the very local volume examined by most
previous studies, thus a direct void by void comparison is ruled out.
A comparison between voids found in all of these different samples was
made in HV02. Here we concentrate on comparing the properties of
2dFGRS voids with those found in the UZC and PSCz samples using the
same technique.

A concern in our analysis of the UZC and PSCz survey was that the edge
of the survey restricted the sizes of the voids.  This effect is less
important in the 2dFGRS, thus comparison of the distribution of void
sizes may indicate whether analyses of the shallower catalogs was
biased.  The average effective radius of voids in the UZC and PSCz is
14.9$\pm 1.8$ and 14.6$\pm1.4 h^{-1}$Mpc, respectively. The average
effective radius of voids in the 2dFGRS is $14.89 \pm 2.65 h^{-1}$Mpc
and $15.61 \pm 2.84 h^{-1}$Mpc for the NGP and SGP regions,
respectively.  The void sizes are comparable, suggesting that the
sizes of voids in the UZC and PSCz were not underestimated. The
densities of voids are also comparable. Voids in the 2dFGRS have
average density contrast $\delta \rho /\rho=-0.94\pm0.02$ and
$-0.93\pm0,02$ in the NGP and SGP, respectively. The UZC and PSCz
voids have $\delta\rho/\rho=-0.92\pm0.03$ and $-0.96\pm0.01$.  This
agreement and the quite similar percentages of galaxies that are
classified as void galaxies in both studies indicates that this method
of identifying voids is robust with respect to variation in galaxy
selection and survey geometry.

\section{Void Statistics}
\label{sec:vpf}

\subsection{The Void Probability and Underdensity Probability Functions}

We statistically quantify the 
the probability of
finding voids in the 2dFGRS using the
Void Probability Function (VPF) (White 1979; see also Lachi\`eze-Rey, da
Costa \& Maurogordato 1992; Watson \& Rowan-Robinson 1993; Vogeley,
Geller \& Huchra 1991; Benson et al. 2003). The VPF gives the probability
that a randomly selected volume of a survey contains no galaxies. In
reality the region might contain galaxies that are below the magnitude
limit of the survey but there are no survey galaxies in the randomly
chosen region.

The VPF depends on all the n-point correlation functions according to
(White 1979)
\begin{equation}
P_0(n,V) = {\rm exp} \left [ \sum_{N=1}^{\infty} \frac{(-n)^N}{N!} \int_V w_N({\bf x_1,...,x_N})d^3x_1...d^3x_N \right ]
\end{equation}
where $n$ is the average number density of galaxies, $w_N$ are the n-point
correlation functions and {\bf $x_i$} are the galaxy positions in the
volume $V$. The simplest case is a Poisson distribution where
\begin{equation}
P_0 = e^{-nV}
\end{equation}
Here we examine spherically symmetric test volumes, thus $P_0=P_0(r)$,
where $r$ is the radius of a test sphere.
 
A disadvantage of the VPF is its sensitivity to shot noise, because a
single galaxy changes the contribution of a large void to the
statistic.  A related, somewhat more robust, statistic is the
underdensity probability function (UPF), which measures the probability
of the density 
lying below a selected threshold in a region of scale $r$,
\begin{equation}
U(r) = P(\rho < \rho_{\rm crit}; r)
\end{equation}
which is an integral over the low-$\rho$ tail of the density
distribution function. As with the VPF, we measure the UPF for
spherically symmetric regions. As discussed below, we measure the UPF
for an underdensity threshold of $\delta\rho/\rho=-0.8$, just slightly
denser than the mean underdensity of the large voids detected in
section 4 above.

The method of calculating both the VPF ($P_0(r)$) and UPF ($U(r)$) is
straightforward. A point is randomly chosen within the survey
volume. The first test is to see if a sphere of radius $r$ lies
completely in the survey or not. If so, then the number of galaxies
that lies in this sphere is counted. This is repeated many times for
different sphere radii. The VPF is then simply estimated as $P_0 =
N_0/N_{\rm test}$ and the estimator for the UPF is $U(r) = N_{(\rho <
\rho_{\rm crit})}/N_{\rm test}$.

The probability of finding a void increases with distance in flux
limited samples due to the decreasing selection function as faint
galaxies become too faint to be detected by the survey. Some earlier
analyses (Vogeley, Geller \& Huchra 1991; Lachi\`eze-Rey, da Costa \&
Maurogordato 1992; Watson \& Rowan-Robinson 1993) concentrated on
measuring the VPF from flux-limited samples, to maximize use of
smaller redshift samples.  The sizes of voids were rescaled to
compensate for the decreasing selection function.  Here we follow
Vogeley et al. (1994), who measured the VPF of volume-limited samples
from the CfA2 survey with limiting absolute magnitudes M$_{\rm lim}-5
{\rm log}h=-18.5, -19, -19.5$ and $-20$. The VPF results from
volume-limited samples are easier to interpret as there is no need to
rescale the sizes of voids and one can compare results from samples
with the same average number density drawn from other surveys.

\subsection{VPF Results for the 2dFGRS}

\begin{figure} 
\begin{centering}
\begin{tabular}{cc}
{\epsfxsize=8truecm \epsfysize=8truecm \epsfbox[40 120 600 700]{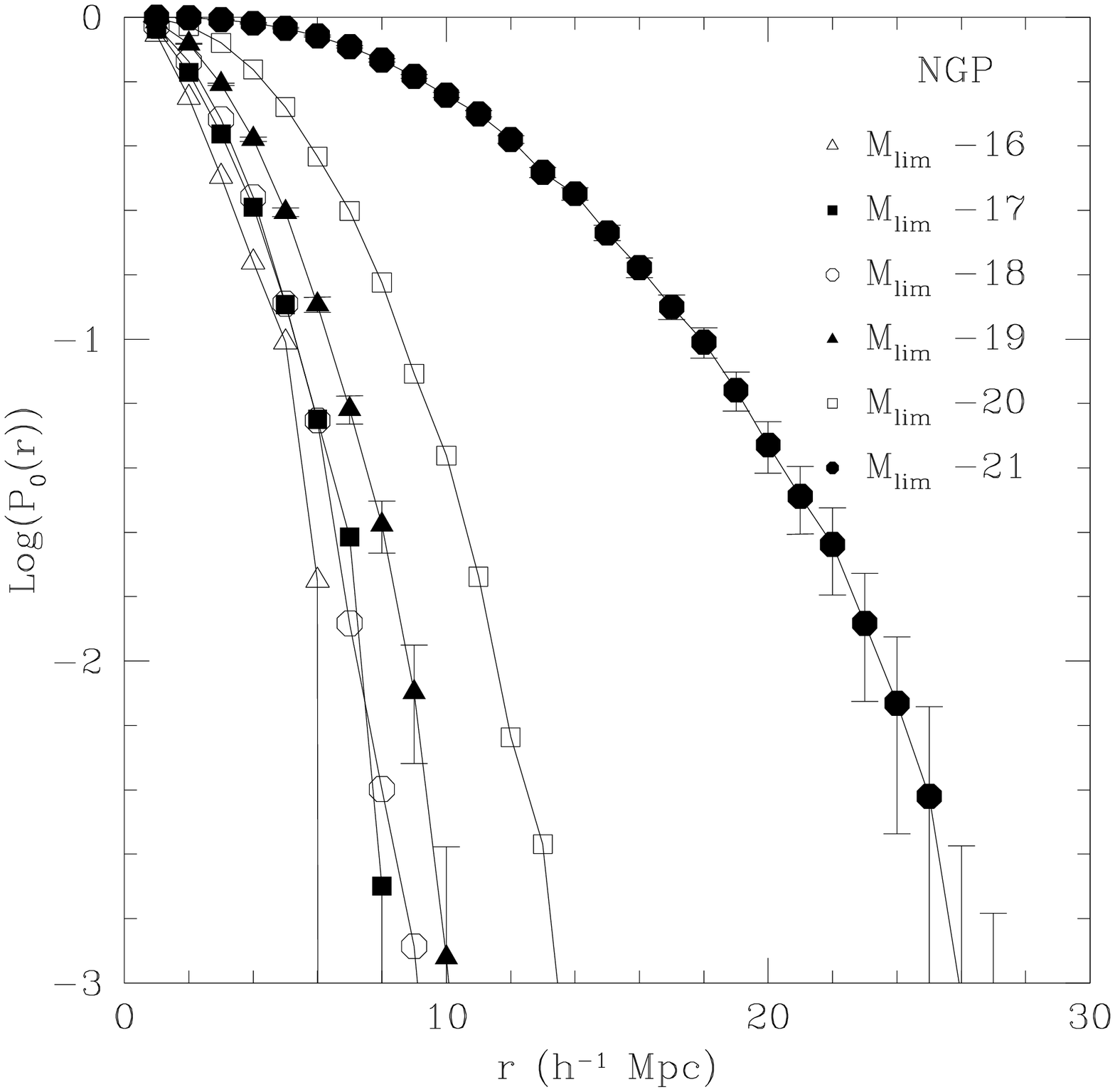}} &
{\epsfxsize=8truecm \epsfysize=8truecm \epsfbox[40 120 600 700]{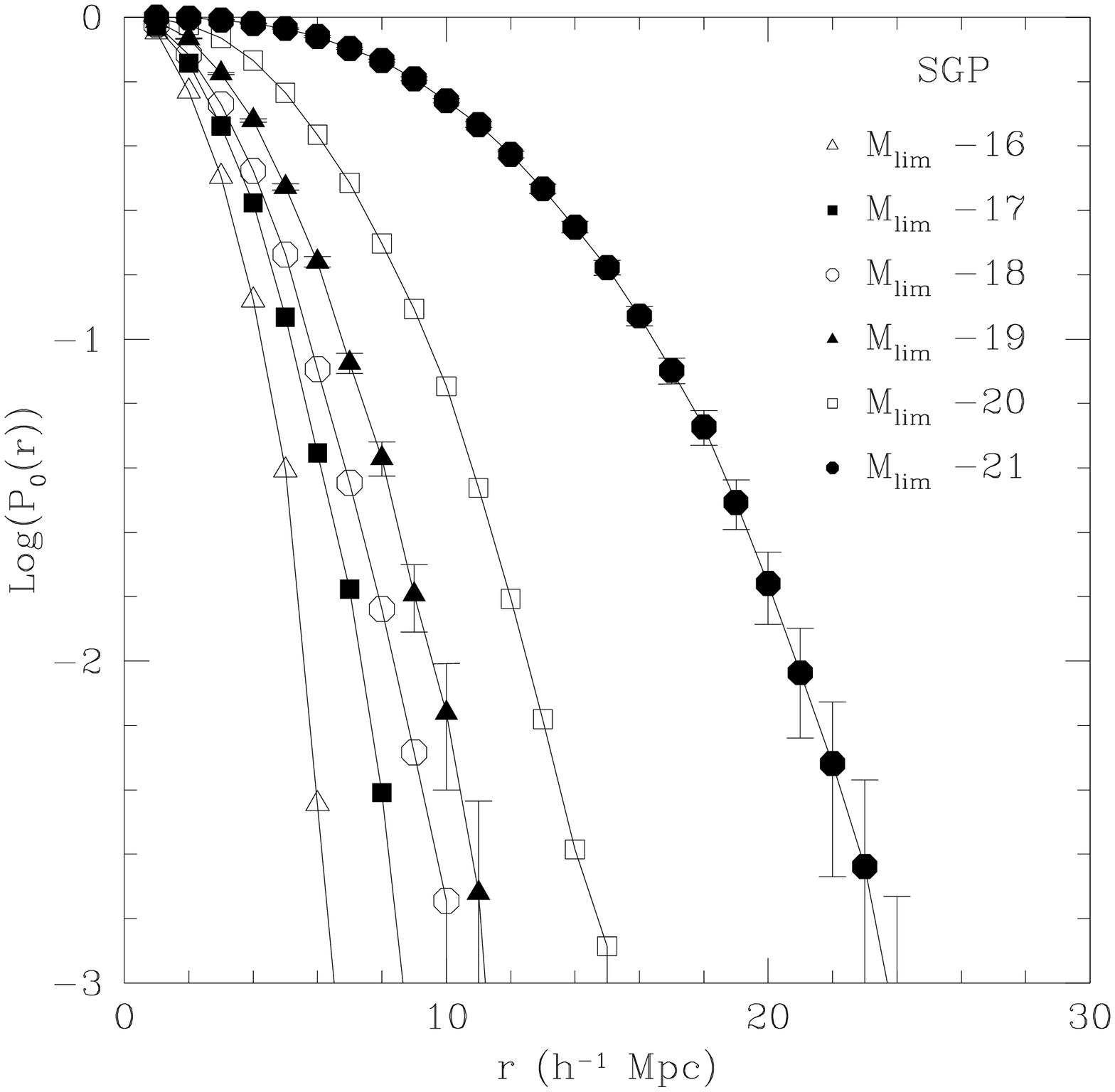}} \\
\end{tabular}
\caption{Void Probability Functions (VPF) for volume-limited samples
extracted from the NGP (left hand plot) and SGP (right hand plot)
regions of the 2dFGRS. The details of each of the samples are given in
Table \ref{tab:volsamps}. Errorbars on the $M_{\rm lim} = -19$ and $-21$
samples are the 1$\sigma$ variation due to the finite number of
independent volumes in the 2dFGRS. }
\label{fig:vpf_ns}
\end{centering}
\end{figure}

\begin{figure}  
\begin{centering}
\begin{tabular}{c}
{\epsfxsize=8truecm \epsfysize=8truecm \epsfbox[40 120 600 700]{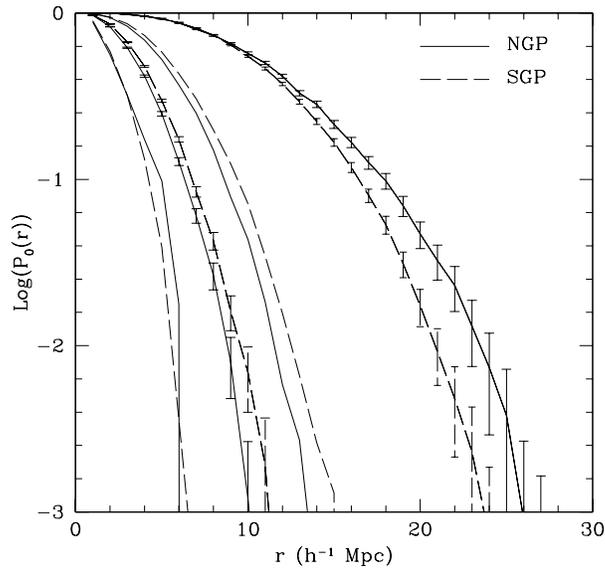}} \\
\end{tabular}
\caption{Comparison of VPF's measured for the NGP (solid lines) and SGP (dashed
lines) of 4 of the 6 samples shown above:  $M_{\rm lim}=-16, -19, -20$ and $-21$
(z$_{\rm max}$=0.039, 0.126, 0.182 and 0.270). We see that locally (for all but the brightest sample),
there is a higher probability of finding voids in the SGP (the lines
are shifted to the right). This is consistent with the results of {\tt
voidfinder}; the SGP is emptier than the NGP. In the brightest
sample, voids are more likely to be detected in the
NGP.}
\label{fig:vpf_comp}
\end{centering}
\end{figure}

We measure the VPF of volume-limited samples drawn from the
2dFGRS. The details of each sample are given in Table
\ref{tab:volsamps}.  We present results for the VPF in Figures
\ref{fig:vpf_ns} and \ref{fig:vpf_comp}. In Figure
\ref{fig:vpf_ns} we show the VPF from various volume limited samples
extracted from the NGP (left hand plot) and SGP (right hand plot). For
clarity, error bars are shown on the M$_{\rm lim}-5 {\rm log}h=-19$
and $-21$ samples.  These uncertainties are the 1$\sigma$ variation
due to the finite number of independent volumes in the 2dFGRS.  The
binomial distribution of void counts yields the uncertainty estimate
\begin{equation}
\sigma(P_0) = \frac{(P_0 - P_0^2)^{1/2}}{N_{\rm indep}^{1/2}}
\label{eq:errors}
\end{equation}
where $N_{\rm indep} = V_{\rm survey}/(4 \pi r_{\rm test}^3 /3)$. As
expected, the void probabilities are larger in samples of brighter
galaxies because the average number density is smaller. We estimate
the VPF for 6 volume-limited samples with limiting absolute magnitudes
M$_{\rm lim}-5 {\rm log}h$ from $-16$ to $-21$. The values of the
VPF are available in Tables \ref{tab:vpfn} and \ref{tab:vpfs}.
This study of void
statistics probes the widest range of galaxy luminosities to date.

\begin{table}
\begin{centering}
\begin{tabular}{cccccc}
M$_{\rm lim}-5 {\rm log}h$ & z$_{\rm max}$ & N$_{\rm gal}$ NGP &
N$_{\rm gal}$ SGP & $\bar{n}$ NGP & $\bar{n}$ SGP \\ 
& & & & $h^{-3}$Mpc$^3$ & $h^{-3}$Mpc$^3$ \\ \hline 
-16 & 0.039 & 4076 & 4928 & 0.0560 & 0.0477 \\ 
-17 & 0.058 & 8686 & 9345 & 0.0360 & 0.0273 \\
-18 & 0.087 & 21128 & 21752 & 0.0282 & 0.0200 \\ 
-19 & 0.126 & 27162 & 33667 & 0.0114 & 0.0099 \\ 
-20 & 0.182 & 19151 & 25139 & 0.0028 & 0.0010 \\ 
-21 & 0.270 & 4123 & 7084 & 0.00019 & 0.00023 \\ \hline
\end{tabular}
\caption{The properties of the volume limited samples used in the VPF and
UPF calculations.}
\label{tab:volsamps}
\end{centering}
\end{table}

We compare the VPF of the NGP (solid lines) and SGP (dashed lines) of
4 of the 6 samples in Figure \ref{fig:vpf_comp}. Only the samples with
M$_{\rm lim}-5 {\rm log}h = -17, -19, -20$ and $-21$ (z$_{\rm
max}$=0.039, 0.126, 0.182 and 0.270) are shown for clarity. We see
that locally, i.e. in the fainter sub-samples, there is a higher
probability of finding voids in the SGP than in the the NGP (the lines
are shifted to the right).  This result is consistent with the result
discussed above using {\tt voidfinder}, which shows that the SGP is
somewhat emptier than the NGP.  In the brightest sample, however, we
find that voids are more likely to be detected in the NGP.  We
attribute these variations to large-scale inhomogeneity on the scales
of these samples. It can be seen in Table \ref{tab:volsamps} that in
the all but the brightest samples, the SGP has a lower number density
of galaxies than the NGP, thus a higher probability of detecting voids
would be expected. In the M$_{\rm lim}-5 {\rm log}h=-21$ samples, the
NGP has a slightly lower number density than the SGP and the chance of
finding voids is larger.

\begin{figure} 
\begin{centering}
\begin{tabular}{c}
{\epsfxsize=8truecm \epsfysize=8truecm \epsfbox[40 120 600 700]{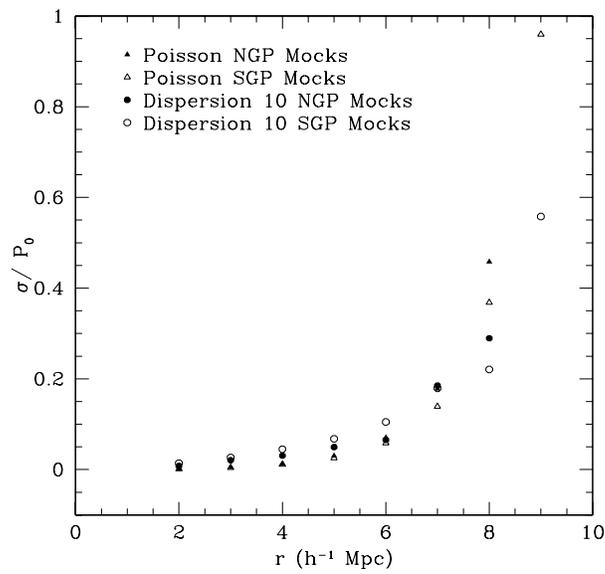}} \\
\end{tabular}
\caption{Comparison of methods for estimating uncertainties in the VPF,
using the Poisson type errors of equation \ref{eq:errors} (triangles)
and the dispersion over 10 mock catalogs (circles). We show the
results for the NGP (filled symbols) and the SGP (open symbols). The
two methods give similar errors, although the Poisson errors are
smaller on small scales and larger on large scales, as compared to the
errors from the mock catalogs. }
\label{fig:ers}
\end{centering}
\end{figure}

We use mock catalogs from the Hubble Volume (Evrard et al. 2002)
simulation to assess the different ways in which errors of the VPF can
be estimated.  In Figure \ref{fig:ers}, we compare the errors found
using equation \ref{eq:errors} (triangles), which are referred to as
Poisson errors as they depend on the number of independent volumes in
the survey area, and the dispersion over 10 mock catalogs
(circles). We show the results for the NGP (filled symbols) and SGP
(open symbols).  The errors follow the same trend and are in
reasonable agreement.  On small scales, the Poisson errors
underestimate the error from the mocks, whereas on large scales, the
Poisson errors overestimate the errors from the mock catalogs, as the
number of independent volumes decreases as the size of the void grows.

\subsection{Underdensity Probabilities of the 2dFGRS}

\begin{figure} 
\begin{centering}
\begin{tabular}{cc}
{\epsfxsize=8truecm \epsfysize=8truecm \epsfbox[40 120 600 700]{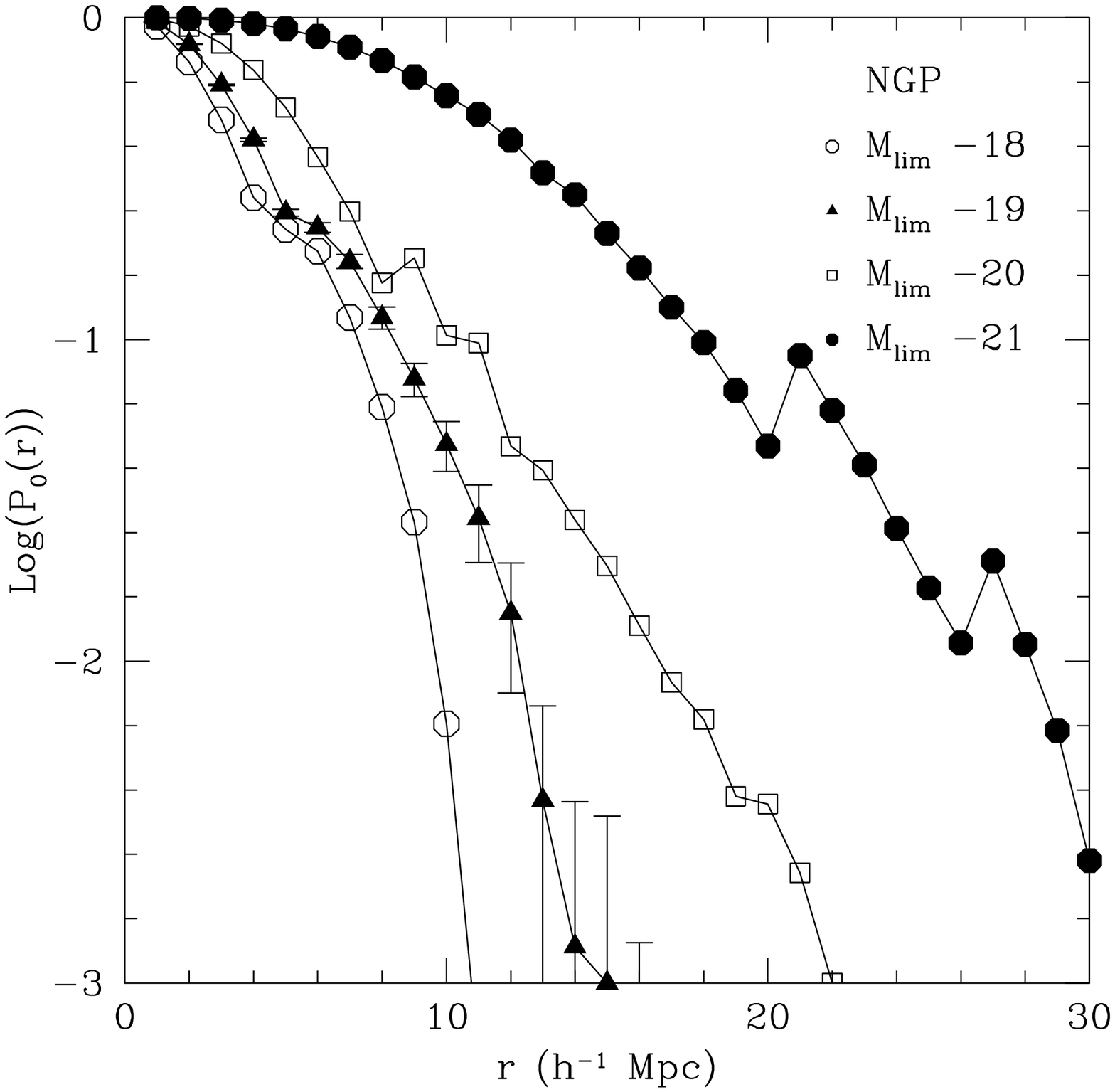}} &
{\epsfxsize=8truecm \epsfysize=8truecm \epsfbox[40 120 600 700]{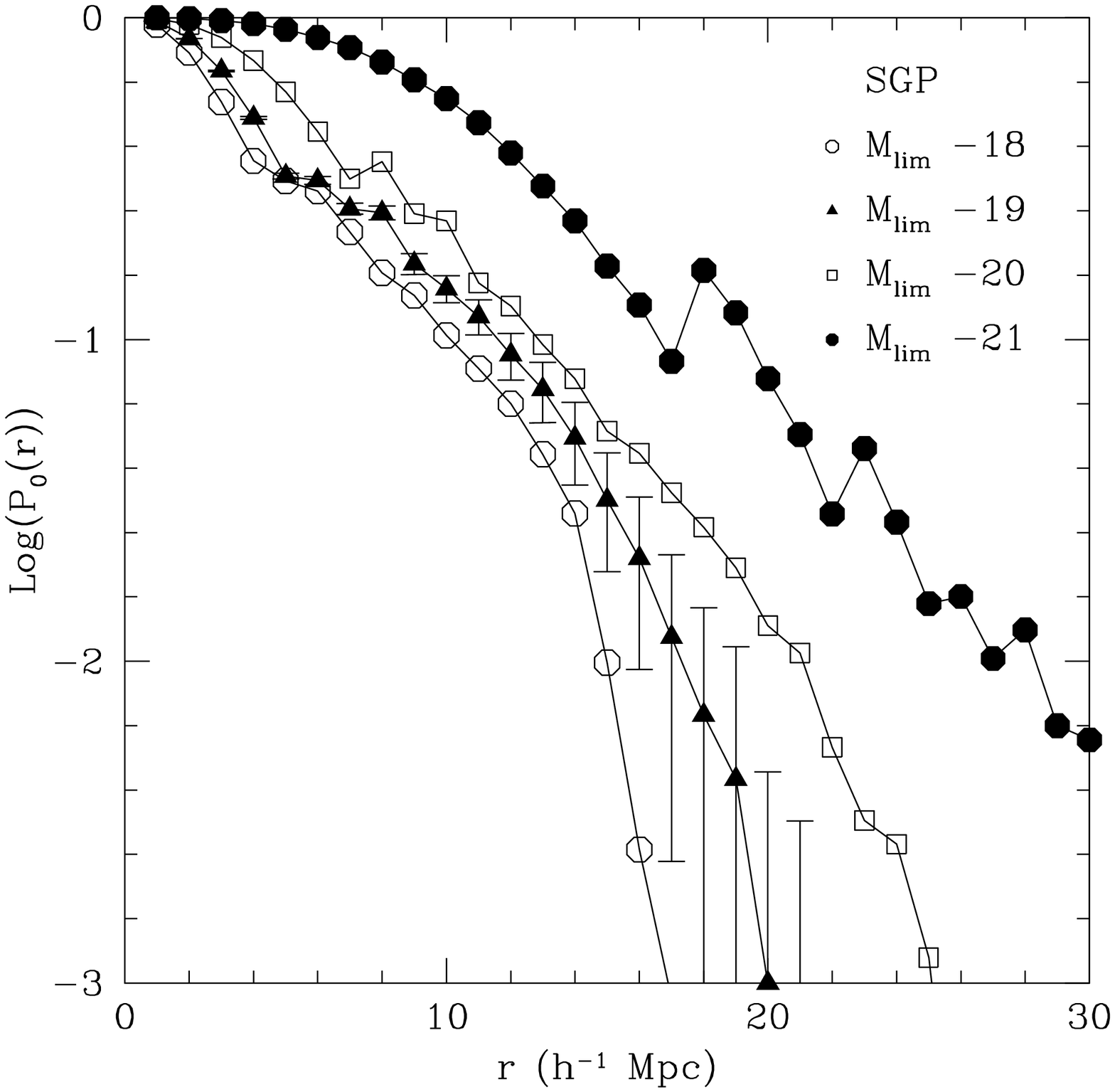}} \\
\end{tabular}
\caption{Underdensity Probability Function at density threshold
$\delta\rho/\rho=-0.8$ for volume-limited samples extracted from the
NGP (left hand plot) and SGP (right hand plot). Details of each of the
samples are given in Table \ref{tab:volsamps}. For clarity, error bars
are shown on the $-19$ samples only and are the 1$\sigma$ variation
due to the finite number of independent volumes in the 2dFGRS.}
\label{fig:upf_ns}
\end{centering}
\end{figure}

In Figure \ref{fig:upf_ns} we present results of the UPF estimated for
the same volume-limited samples of the NGP and SGP regions.  For this
analysis, we choose an underdensity threshold of
$\delta\rho/\rho=-0.8$, which is slightly less restrictive than the
density contrast $\delta\rho/\rho<-0.9$ of the the voids found in
section \ref{sec:res}.  If we were to choose a more extreme density
threshold, the UPF would be almost equivalent to the VPF because voids
are discrete objects.  As estimated here, the UPF probes the regions
near the boundary between the voids and denser structures.

In Figure \ref{fig:upf_ns} we plot the UPF for a threshold
$\delta\rho/\rho=-0.8$ for volume-limited samples with with M$_{\rm
lim}-5 {\rm log}h=-18, -19, -20$ and $-21$ and the data are
available in Tables \ref{tab:upfn} and \ref{tab:upfs}. As for the VPF, the
chance of finding an underdense region increases as the magnitude
limit of the samples brightens and the number density of objects in
the sample decreases.

\begin{figure} 
\begin{centering}
\begin{tabular}{c}
{\epsfxsize=8truecm \epsfysize=8truecm \epsfbox[40 120 600 700]{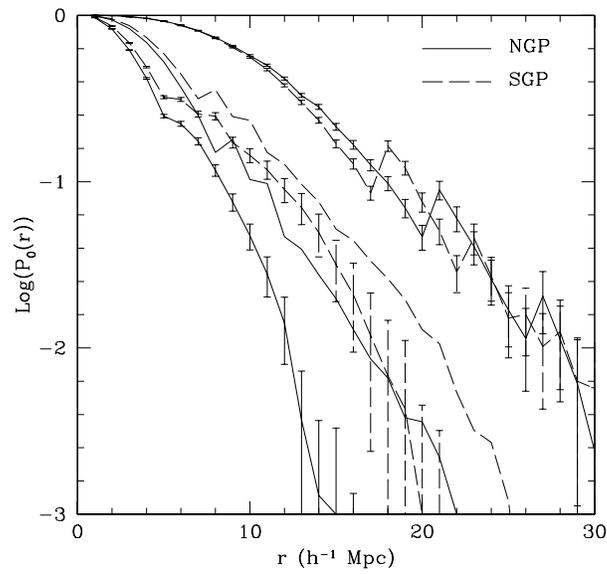}} \\
\end{tabular}
\caption{Comparison of the UPF's of the NGP (solid lines) and
SGP (dashed lines) the samples with $ M_{\rm lim}=-19, -20$ and $-21$
(z$_{\rm max}$=0.126, 0.182 and 0.270). Again, we see that locally,
there is a higher probability of finding underdense regions in the
SGP, the lines are shifted to the right. In this case though, there is
good agreement between the brightest samples.}
\label{fig:upf_comp}
\end{centering}
\end{figure}

In Figure \ref{fig:upf_comp}, we compare the UPF's for three
volume-limited samples with $M_{\rm lim}-5 {\rm log}h=-19, -20$ and
$-21$ of both the NGP and SGP regions.  Again, we find that the SGP is
more underdense than the NGP, but in this case we find good agreement
in the brightest samples.

\subsection{Comparison With Previous Work}

\begin{figure} 
\begin{centering}
\begin{tabular}{c}
{\epsfxsize=8truecm \epsfysize=8truecm \epsfbox[40 120 600 700]{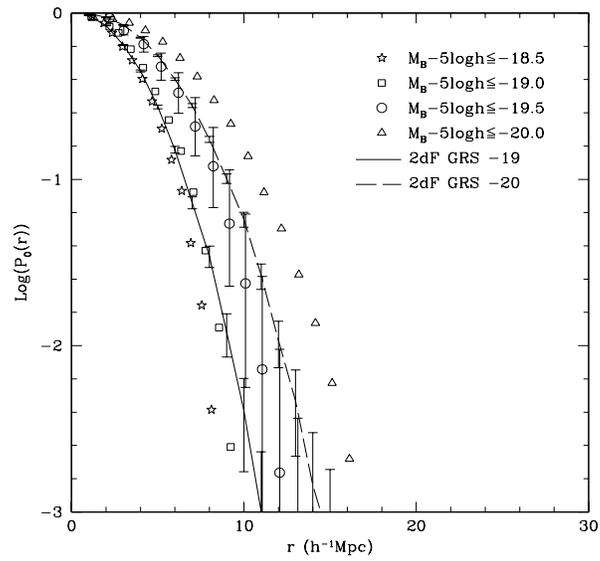}} \\
\end{tabular}
\caption{Comparison of the 2dFGRS VPF's with earlier results from the
CfA2 (Vogeley et al. 1994). Symbols show the VPF's from CfA2 and lines
show results from the 2dFGRS. Errorbars are shown on the $-19.5$ CfA2
sample and on the 2dFGRS samples.  Taking into account a zeropoint
shift of roughly $m_Z=b_j+0.5$, we note that the $-18.5$ CfA2 sample
and $-19.0$ 2dFGRS samples agree within the errors (which are strongly
correlated), as do the $-19.5$ CfA2 and $-20.0$ 2dFGRS samples.  }
\label{fig:cfa}
\end{centering}
\end{figure}

The VPF and UPF have previously been estimated from the Center for
Astrophysics survey (de Lapparent et al. 1986; Geller \& Huchra 1989;
Huchra et al. 1999) by Vogeley et al. (1994).  In Figure
\ref{fig:cfa}, we compare the measurements of the VPF of the CfA
survey to those from the 2dFGRS. To get an overall measurement for the
2dFGRS for each magnitude cut, we average the VPF's of the NGP and
SGP, weighted by their respective errors.  Uncertainties on the VPF's
of the 2dFGRS samples are significantly smaller than those for the CfA
due to the larger volume of the 2dFGRS.  

Comparison of these results is complicated by the possibility,
suggested by comparison with CCD photometry, that the Zwicky magnitude
system is not simply offset in zeropoint from $b_J$, but also suffers
a significant scale error (Gazta\~naga \& Dalton 2000). Here we make
the simplest assumption -- assuming an offset of $m_Z=b_J+0.5$ (as
suggested by Efstathiou et al. 1988 and Lin et al. 1994) -- that we
can compare the VPF's of the $M_{\rm lim}-5 {\rm log}h=-19.0$ 2dFGRS
sample and $-18.5$ CfA2 sample. These VPF's agree within the quite
large uncertainties of the CfA2 sample.  The VPF of the latter is
systematically smaller, but note carefully that values of $P_0$ at
different $r$ are not independent. Although there is better agreement
if we compare the $-19.0$ samples, it seems more likely that the
smaller volume of CfA2 causes the VPF to be low than that the
magnitude systems match. Likewise, we may compare the $-20.0$ 2dFGRS
sample with the $-19.5$ CfA2 sample. Again, these agree within the
large uncertainties and again we suspect that a combination of cosmic
variance and problems with comparing the magnitude systems causes the
offset.

\subsection{Comparison with VPF's in Simulations}

\begin{figure} 
\begin{centering}
\begin{tabular}{c}
{\epsfxsize=8truecm \epsfysize=8truecm \epsfbox[40 120 600 700]{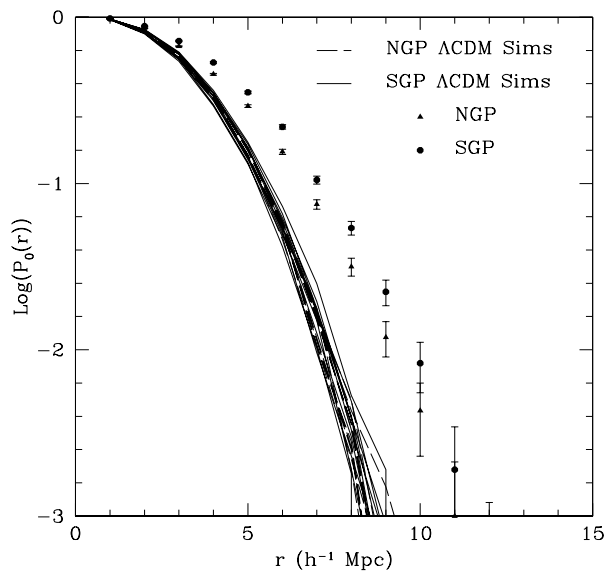}} \\
\end{tabular}
\caption{Comparison of the VPF of the 2dFGRS with the VPF of
mock catalogs drawn from a dark matter only $\Lambda$CDM
simulation, the {\it Hubble Volume}. The dark matter only mock
catalogs contain too much substructure and fail to match the VPF's of
the data, as seen in Benson et al. (2003). }
\label{fig:vpf_sims}
\end{centering}
\end{figure}

In Figure \ref{fig:vpf_sims}, we compare the VPF's of the data with
the VPF of mock catalogs drawn from a dark matter only $\Lambda$CDM
simulation, the {\it Hubble Volume} (Evrard et al. 2002). The dark
matter only mock catalogs contain too much substructure on small
scales and fail to match the VPF's of the data, as seen in Benson et
al. (2003). 

\begin{figure} 
\begin{centering}
\begin{tabular}{c}
{\epsfxsize=8truecm \epsfysize=8truecm \epsfbox[40 120 600 700]{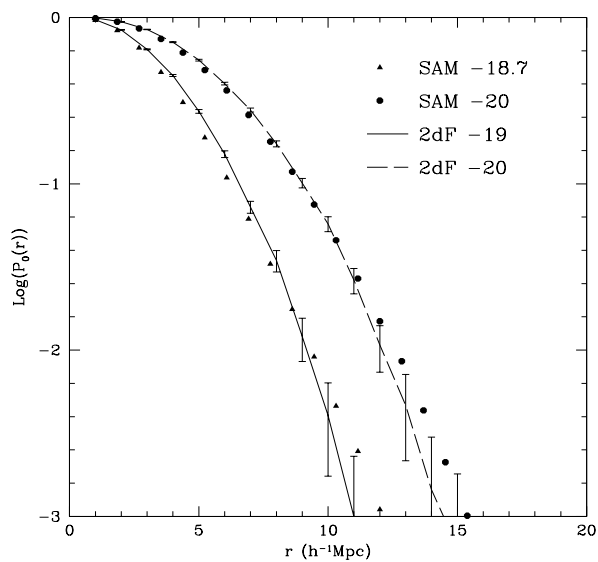}} \\
\end{tabular}
\caption{Comparison of the 2dFGRS VPF's and those from the mock
semi-analytic galaxy catalogs of Benson et al. (2003). There is
excellent agreement between the theory and the data.}
\label{fig:sam}
\end{centering}
\end{figure}

A better comparison between theory and data can be made by using mock
galaxy catalogs constructed using semi-analytic models of galaxy
formation, as done by Benson et al. (2003), who measure the VPF of
mock 2dFGRS samples.  In Figure \ref{fig:sam}, we show the VPF's from
the models (symbols) and 2dFGRS data (lines). The models have $M_{\rm
lim}-5 {\rm log}h=-18.7$ and $-20$, close to the limits of $-19$ and
$-20$ for our observed 2dFGRS samples.  The agreement between the data
and theory is excellent. This agreement, in contrast to the strong
disagreement with the VPF of random samples of the dark matter
distribution, shows the importance of including more detailed physics
in the prescription for forming galaxies in simulations. We look
forward to comparisons of these measurements of void statistics with
results from high-resolution cosmological hydrodynamic/N-body
simulations.

\section{Summary and Conclusions}
\label{sec:conc}


We quantify the distribution of voids in the 2dFGRS by applying an
objective void-finding algorithm ({\tt voidfinder}) and by measuring
the frequency of empty and underdense regions. These statistics
provide strong constraints on models for structure formation.


Results of this work include a new catalog of 289 voids: 116 in the
NGP region and 173 in the SGP region. This sample is almost a factor
of 10 larger than previously identified. 
The larger number of voids in the SGP reflects both its larger volume and slightly lower average density of galaxies.
These voids have minimum
radii of $10h^{-1}$Mpc, an average effective radius of $15h^{-1}$Mpc,
and average density contrast $\delta\rho/\rho=-0.94$.

Examination of the density profile of the voids shows that the {\tt
voidfinder} algorithm identifies the inner region of the voids where
the galaxy density is nearly constant, varying from
$\delta\rho/\rho\sim -0.9$ at the outer edge and decreasing toward
zero density near the center.  Just outside the boundary of the voids,
the galaxy density sharply rises, although the mean density remains
well below $\delta\rho/\rho=-0.5$ out to 150\% of the void radius.

These extremely underdense voids fill 40\% of the volume of the
survey. The voids contain 5\% of all galaxies in the sample.  All of
the measured properties of voids found in the 2dFGRS agree with void
properties of the more nearby galaxy distributions sampled by 
the UZC and PSCz survey (HV02),
indicating that this algorithm for identifying voids is robust with
respect to variation in galaxy selection and survey geometry.
Comparable results are found using a similar algorithm applied to the
IRAS 1.2Jy (El-Ad, Piran, \& da Costa 1997), SSRS2 (El-Ad \& Piran
1997), and ORS (El-Ad \& Piran 2000).


We measure the Void Probability Function and Underdensity Probability
Function of volume-limited samples. These statistics quantify the
frequency of completely empty and underdense spheres, respectively.
We present the VPF's and UPF's for the widest range of absolute
magnitudes examined to date: from $-16$ to $-21$ for the VPF and $-18$
to $-21$ for the UPF.  Variation of these statistics between the two
survey regions reflect underdensity of the SGP relative to the NGP out
to comoving distance 400$h^{-1}$Mpc.  We find reasonable agreement
between the VPF's of the 2dFGRS and results from the largest previous
study, of the CfA2 survey (Vogeley et al. 1994).  The larger volume of
the 2dFGRS makes these new VPF measurements significantly more precise
than could be estimated from CfA2.

Comparison of void statistics of 2dFGRS with results for the
distribution of dark matter particles in large cosmological N-body
simulations show that the matter distribution of the models does not
have the correct VPF. However, application of semi-analytic modeling
to results of high-resolution dark matter simulations (Benson et
al. 2003) yields mock galaxy catalogs that provide uncanny agreement
with the VPF measured for the 2dFGRS. 

Void statistics will yield more precise constraints on structure
formation models as we examine samples that cover even larger volume
and include more information about intrinsic properties of the
galaxies. The larger volume, five-band photometry, and
medium-resolution spectroscopy of the Sloan Digital Sky Survey will
allow precise measurement of void statistics, even after subdividing
samples by galaxy property. On the theoretical side, improvement in
the dynamic range of cosmological simulations and the inclusion of
more sophisticated treatment of hydrodynamics and energy feedback will
yield more realistic modeling of the formation of the few galaxies
that lie in cosmic voids.

\acknowledgments MSV acknowledges support from NSF grant
AST-0071201. We acknowledge the enormous efforts of all involved with
the 2dF project and thank them for providing this wonderful data set.

\begin{table}
\begin{centering}
\begin{tabular}{ccccccc}
r ($h^{-1}$Mpc) &  -17 & -18 &  
-19 & -20 &  -21  \\ \hline
 1  & 0.9200$\pm$0.0011 & 0.9410$\pm$0.0005 & 0.9680$\pm$0.0002 & 0.9920$\pm$0.0001 & 0.9990$\pm$0.0000  \\
 2  & 0.6740$\pm$0.0055 & 0.7280$\pm$0.0029 & 0.8280$\pm$0.0014 & 0.9390$\pm$0.0005 & 0.9960$\pm$0.0001  \\
 3  & 0.4340$\pm$0.0107 & 0.4820$\pm$0.0059 & 0.6190$\pm$0.0033 & 0.8320$\pm$0.0015 & 0.9830$\pm$0.0003  \\
 4  & 0.2570$\pm$0.0146 & 0.2760$\pm$0.0082 & 0.4180$\pm$0.0052 & 0.6880$\pm$0.0029 & 0.9580$\pm$0.0007  \\
 5  & 0.1280$\pm$0.0156 & 0.1290$\pm$0.0086 & 0.2480$\pm$0.0064 & 0.5260$\pm$0.0043 & 0.9250$\pm$0.0013  \\
 6  & 0.0564$\pm$0.0141 & 0.0559$\pm$0.0077 & 0.1280$\pm$0.0065 & 0.3690$\pm$0.0055 & 0.8740$\pm$0.0022  \\
 7  & 0.0243$\pm$0.0119 & 0.0131$\pm$0.0048 & 0.0606$\pm$0.0059 & 0.2500$\pm$0.0062 & 0.8100$\pm$0.0032  \\
 8  & 0.0020$\pm$0.0042 & 0.0040$\pm$0.0033 & 0.0265$\pm$0.0048 & 0.1500$\pm$0.0063 & 0.7350$\pm$0.0044  \\
 9  & 0.0000$\pm$0.0000 & 0.0013$\pm$0.0022 & 0.0080$\pm$0.0032 & 0.0781$\pm$0.0056 & 0.6550$\pm$0.0057  \\
10  & 0.0000$\pm$0.0000 & 0.0001$\pm$0.0007 & 0.0012$\pm$0.0015 & 0.0435$\pm$0.0050 & 0.5730$\pm$0.0069  \\
11  & 0.0000$\pm$0.0000 & 0.0000$\pm$0.0000 & 0.0001$\pm$0.0005 & 0.0183$\pm$0.0038 & 0.5010$\pm$0.0081  \\
12  & 0.0000$\pm$0.0000 & 0.0000$\pm$0.0000 & 0.0000$\pm$0.0000 & 0.0058$\pm$0.0025 & 0.4170$\pm$0.0091  \\
13  & 0.0000$\pm$0.0000 & 0.0000$\pm$0.0000 & 0.0000$\pm$0.0000 & 0.0027$\pm$0.0019 & 0.3300$\pm$0.0098  \\
14  & 0.0000$\pm$0.0000 & 0.0000$\pm$0.0000 & 0.0000$\pm$0.0000 & 0.0003$\pm$0.0007 & 0.2830$\pm$0.0105  \\
15  & 0.0000$\pm$0.0000 & 0.0000$\pm$0.0000 & 0.0000$\pm$0.0000 & 0.0000$\pm$0.0000 & 0.2140$\pm$0.0106  \\
16  & 0.0000$\pm$0.0000 & 0.0000$\pm$0.0000 & 0.0000$\pm$0.0000 & 0.0000$\pm$0.0000 & 0.1670$\pm$0.0106 \\
17  & 0.0000$\pm$0.0000 & 0.0000$\pm$0.0000 & 0.0000$\pm$0.0000 & 0.0000$\pm$0.0000 & 0.1260$\pm$0.0103  \\
18  & 0.0000$\pm$0.0000 & 0.0000$\pm$0.0000 & 0.0000$\pm$0.0000 & 0.0000$\pm$0.0000 & 0.0979$\pm$0.0101  \\
19  & 0.0000$\pm$0.0000 & 0.0000$\pm$0.0000 & 0.0000$\pm$0.0000 & 0.0000$\pm$0.0000 & 0.0694$\pm$0.0093  \\
20  & 0.0000$\pm$0.0000 & 0.0000$\pm$0.0000 & 0.0000$\pm$0.0000 & 0.0000$\pm$0.0000 & 0.0469$\pm$0.0084  \\
21  & 0.0000$\pm$0.0000 & 0.0000$\pm$0.0000 & 0.0000$\pm$0.0000 & 0.0000$\pm$0.0000 & 0.0325$\pm$0.0076  \\
22  & 0.0000$\pm$0.0000 & 0.0000$\pm$0.0000 & 0.0000$\pm$0.0000 & 0.0000$\pm$0.0000 & 0.0230$\pm$0.0069  \\ \hline
\end{tabular}
\caption{Void Probability Function $P_0(r)$ measured measured for volume-limited samples of the NGP region with absolute magnitude
  limits M$_{\rm lim}-5 {\rm log}h=-17, -18, -19, -20$ and
  $-21$.}
\label{tab:vpfn}
\end{centering}
\end{table}

\begin{table}
\begin{centering}
\begin{tabular}{cccccc}
r ($h^{-1}$Mpc) &  -17 &  -18 &  -19 &  -20 &  -21  \\ \hline
1 & 0.9350$\pm$0.0009 & 0.9500$\pm$0.0004 & 0.9730$\pm$0.0002 & 0.9920$\pm$0.0001 & 0.9990$\pm$0.0000  \\
 2 & 0.7200$\pm$0.0044 & 0.7650$\pm$0.0023 & 0.8580$\pm$0.0011 & 0.9450$\pm$0.0004 & 0.9950$\pm$0.0001  \\
 3 & 0.4590$\pm$0.0091 & 0.5370$\pm$0.0050 & 0.6680$\pm$0.0027 & 0.8630$\pm$0.0012 & 0.9800$\pm$0.0003  \\
 4 & 0.2650$\pm$0.0124 & 0.3330$\pm$0.0072 & 0.4780$\pm$0.0044 & 0.7330$\pm$0.0023 & 0.9560$\pm$0.0006  \\
 5 & 0.1170$\pm$0.0126 & 0.1830$\pm$0.0083 & 0.2970$\pm$0.0057 & 0.5830$\pm$0.0036 & 0.9220$\pm$0.0011  \\
 6 & 0.0443$\pm$0.0106 & 0.0807$\pm$0.0077 & 0.1740$\pm$0.0062 & 0.4320$\pm$0.0047 & 0.8710$\pm$0.0018  \\
 7 & 0.0167$\pm$0.0083 & 0.0358$\pm$0.0066 & 0.0843$\pm$0.0057 & 0.3060$\pm$0.0056 & 0.7990$\pm$0.0028  \\
 8 & 0.0039$\pm$0.0049 & 0.0145$\pm$0.0052 & 0.0427$\pm$0.0051 & 0.1980$\pm$0.0059 & 0.7340$\pm$0.0037  \\
 9 & 0.0005$\pm$0.0021 & 0.0052$\pm$0.0037 & 0.0161$\pm$0.0038 & 0.1240$\pm$0.0058 & 0.6440$\pm$0.0048  \\
10 & 0.0000$\pm$0.0000 & 0.0018$\pm$0.0026 & 0.0069$\pm$0.0029 & 0.0714$\pm$0.0053 & 0.5500$\pm$0.0059  \\
11 & 0.0000$\pm$0.0000 & 0.0000$\pm$0.0000 & 0.0019$\pm$0.0018 & 0.0345$\pm$0.0043 & 0.4640$\pm$0.0068  \\
12 & 0.0000$\pm$0.0000 & 0.0000$\pm$0.0000 & 0.0001$\pm$0.0005 & 0.0156$\pm$0.0034 & 0.3750$\pm$0.0075  \\
13 & 0.0000$\pm$0.0000 & 0.0000$\pm$0.0000 & 0.0000$\pm$0.0000 & 0.0066$\pm$0.0025 & 0.2930$\pm$0.0079  \\
14 & 0.0000$\pm$0.0000 & 0.0000$\pm$0.0000 & 0.0000$\pm$0.0000 & 0.0026$\pm$0.0017 & 0.2230$\pm$0.0081  \\
15 & 0.0000$\pm$0.0000 & 0.0000$\pm$0.0000 & 0.0000$\pm$0.0000 & 0.0013$\pm$0.0014 & 0.1670$\pm$0.0081  \\
16 & 0.0000$\pm$0.0000 & 0.0000$\pm$0.0000 & 0.0000$\pm$0.0000 & 0.0000$\pm$0.0000 & 0.1180$\pm$0.0077  \\
17 & 0.0000$\pm$0.0000 & 0.0000$\pm$0.0000 & 0.0000$\pm$0.0000 & 0.0000$\pm$0.0000 & 0.0800$\pm$0.0071  \\
18 & 0.0000$\pm$0.0000 & 0.0000$\pm$0.0000 & 0.0000$\pm$0.0000 & 0.0000$\pm$0.0000 & 0.0534$\pm$0.0064  \\
19 & 0.0000$\pm$0.0000 & 0.0000$\pm$0.0000 & 0.0000$\pm$0.0000 & 0.0000$\pm$0.0000 & 0.0311$\pm$0.0054  \\
20 & 0.0000$\pm$0.0000 & 0.0000$\pm$0.0000 & 0.0000$\pm$0.0000 & 0.0000$\pm$0.0000 & 0.0174$\pm$0.0044  \\
21 & 0.0000$\pm$0.0000 & 0.0000$\pm$0.0000 & 0.0000$\pm$0.0000 & 0.0000$\pm$0.0000 & 0.0092$\pm$0.0034  \\
22 & 0.0000$\pm$0.0000 & 0.0000$\pm$0.0000 & 0.0000$\pm$0.0000 & 0.0000$\pm$0.0000 & 0.0048$\pm$0.0027  \\ \hline
\end{tabular}
\caption{Void Probability Function $P_0(r)$ measured for
  volume-limited samples of the SGP region with absolute magnitude
  limits M$_{\rm lim}-5 {\rm log}h=-17, -18, -19, -20$ and $-21$.  }
\label{tab:vpfs}
\end{centering}
\end{table}

\begin{table}
\begin{centering}
\begin{tabular}{ccccc}
r ($h^{-1}$Mpc)  &  -18 &  -19 &  -20 &  -21  \\ \hline
1 & 0.9410$\pm$0.0005 & 0.9680$\pm$0.0002 & 0.9920$\pm$0.0001 & 0.9990$\pm$0.0000  \\
 2 & 0.7290$\pm$0.0029 & 0.8280$\pm$0.0014 & 0.9390$\pm$0.0005 & 0.9960$\pm$0.0001  \\
 3 & 0.4820$\pm$0.0059 & 0.6190$\pm$0.0033 & 0.8320$\pm$0.0015 & 0.9830$\pm$0.0003  \\
 4 & 0.2760$\pm$0.0082 & 0.4180$\pm$0.0052 & 0.6880$\pm$0.0029 & 0.9580$\pm$0.0007  \\
 5 & 0.2200$\pm$0.0106 & 0.2480$\pm$0.0064 & 0.5260$\pm$0.0043 & 0.9250$\pm$0.0013  \\
 6 & 0.1880$\pm$0.0131 & 0.2230$\pm$0.0081 & 0.3690$\pm$0.0055 & 0.8740$\pm$0.0022  \\
 7 & 0.1170$\pm$0.0136 & 0.1750$\pm$0.0093 & 0.2500$\pm$0.0062 & 0.8100$\pm$0.0032  \\
 8 & 0.0617$\pm$0.0124 & 0.1170$\pm$0.0096 & 0.1500$\pm$0.0063 & 0.7360$\pm$0.0044  \\
 9 & 0.0271$\pm$0.0100 & 0.0755$\pm$0.0095 & 0.1790$\pm$0.0080 & 0.6550$\pm$0.0057  \\
10 & 0.0064$\pm$0.0058 & 0.0473$\pm$0.0089 & 0.1030$\pm$0.0075 & 0.5730$\pm$0.0069  \\
11 & 0.0006$\pm$0.0020 & 0.0278$\pm$0.0080 & 0.0974$\pm$0.0084 & 0.5010$\pm$0.0081  \\
12 & 0.0002$\pm$0.0013 & 0.0141$\pm$0.0065 & 0.0466$\pm$0.0068 & 0.4170$\pm$0.0091  \\
13 & 0.0000$\pm$0.0000 & 0.0037$\pm$0.0038 & 0.0392$\pm$0.0071 & 0.3300$\pm$0.0098  \\
14 & 0.0000$\pm$0.0000 & 0.0013$\pm$0.0025 & 0.0275$\pm$0.0067 & 0.2820$\pm$0.0105  \\
15 & 0.0000$\pm$0.0000 & 0.0010$\pm$0.0024 & 0.0198$\pm$0.0063 & 0.2140$\pm$0.0106  \\
16 & 0.0000$\pm$0.0000 & 0.0002$\pm$0.0012 & 0.0129$\pm$0.0056 & 0.1670$\pm$0.0106  \\
17 & 0.0000$\pm$0.0000 & 0.0000$\pm$0.0000 & 0.0086$\pm$0.0050 & 0.1260$\pm$0.0103  \\
18 & 0.0000$\pm$0.0000 & 0.0000$\pm$0.0000 & 0.0066$\pm$0.0048 & 0.0978$\pm$0.0101  \\
19 & 0.0000$\pm$0.0000 & 0.0000$\pm$0.0000 & 0.0038$\pm$0.0040 & 0.0695$\pm$0.0093  \\
20 & 0.0000$\pm$0.0000 & 0.0000$\pm$0.0000 & 0.0036$\pm$0.0042 & 0.0467$\pm$0.0084  \\
21 & 0.0000$\pm$0.0000 & 0.0000$\pm$0.0000 & 0.0022$\pm$0.0035 & 0.0891$\pm$0.0122  \\
22 & 0.0000$\pm$0.0000 & 0.0000$\pm$0.0000 & 0.0010$\pm$0.0025 & 0.0602$\pm$0.0109  \\
23 & 0.0000$\pm$0.0000 & 0.0000$\pm$0.0000 & 0.0001$\pm$0.0009 & 0.0408$\pm$0.0097  \\
24 & 0.0000$\pm$0.0000 & 0.0000$\pm$0.0000 & 0.0000$\pm$0.0000 & 0.0259$\pm$0.0083  \\
25 & 0.0000$\pm$0.0000 & 0.0000$\pm$0.0000 & 0.0000$\pm$0.0000 & 0.0169$\pm$0.0072  \\
26 & 0.0000$\pm$0.0000 & 0.0000$\pm$0.0000 & 0.0000$\pm$0.0000 & 0.0114$\pm$0.0062  \\
27 & 0.0000$\pm$0.0000 & 0.0000$\pm$0.0000 & 0.0000$\pm$0.0000 & 0.0205$\pm$0.0088  \\
28 & 0.0000$\pm$0.0000 & 0.0000$\pm$0.0000 & 0.0000$\pm$0.0000 & 0.0113$\pm$0.0070  \\
29 & 0.0000$\pm$0.0000 & 0.0000$\pm$0.0000 & 0.0000$\pm$0.0000 & 0.0061$\pm$0.0054  \\
30 & 0.0000$\pm$0.0000 & 0.0000$\pm$0.0000 & 0.0000$\pm$0.0000 & 0.0024$\pm$0.0036  \\ \hline
\end{tabular}
\caption{Underdensity Probability Function $U(r)$ (using density threshold
  $\delta\rho/\rho=-0.8$) measured for volume-limited samples of the
  NGP region with absolute magnitude limits M$_{\rm lim}-5 {\rm
    log}h=-18, -19, -20$ and $-21$.}
\label{tab:upfn}
\end{centering}
\end{table}

\begin{table}
\begin{centering}
\begin{tabular}{ccccc}
r ($h^{-1}$Mpc)  &  -18 &  -19 &  -20 &  -21  \\ \hline
1 & 0.9530$\pm$0.0004 & 0.9760$\pm$0.0002 & 0.9930$\pm$0.0001 & 0.9990$\pm$0.0000  \\
 2 & 0.7790$\pm$0.0022 & 0.8620$\pm$0.0011 & 0.9520$\pm$0.0004 & 0.9950$\pm$0.0001  \\
 3 & 0.5470$\pm$0.0050 & 0.6840$\pm$0.0027 & 0.8680$\pm$0.0011 & 0.9790$\pm$0.0003  \\
 4 & 0.3590$\pm$0.0074 & 0.4880$\pm$0.0044 & 0.7360$\pm$0.0023 & 0.9580$\pm$0.0006  \\
 5 & 0.3110$\pm$0.0099 & 0.3220$\pm$0.0058 & 0.5890$\pm$0.0036 & 0.9210$\pm$0.0011  \\
 6 & 0.2890$\pm$0.0128 & 0.3120$\pm$0.0076 & 0.4430$\pm$0.0048 & 0.8680$\pm$0.0019  \\
 7 & 0.2160$\pm$0.0146 & 0.2550$\pm$0.0090 & 0.3160$\pm$0.0056 & 0.8080$\pm$0.0027  \\
 8 & 0.1610$\pm$0.0159 & 0.2480$\pm$0.0109 & 0.3570$\pm$0.0071 & 0.7290$\pm$0.0037  \\
 9 & 0.1370$\pm$0.0178 & 0.1720$\pm$0.0113 & 0.2460$\pm$0.0076 & 0.6420$\pm$0.0048  \\
10 & 0.1030$\pm$0.0184 & 0.1440$\pm$0.0123 & 0.2340$\pm$0.0087 & 0.5600$\pm$0.0058  \\
11 & 0.0813$\pm$0.0191 & 0.1180$\pm$0.0131 & 0.1500$\pm$0.0085 & 0.4720$\pm$0.0068  \\
12 & 0.0631$\pm$0.0194 & 0.0897$\pm$0.0132 & 0.1270$\pm$0.0090 & 0.3800$\pm$0.0075  \\
13 & 0.0440$\pm$0.0184 & 0.0701$\pm$0.0133 & 0.0964$\pm$0.0090 & 0.3000$\pm$0.0080  \\
14 & 0.0288$\pm$0.0168 & 0.0495$\pm$0.0126 & 0.0756$\pm$0.0090 & 0.2340$\pm$0.0083  \\
15 & 0.0099$\pm$0.0110 & 0.0317$\pm$0.0113 & 0.0519$\pm$0.0084 & 0.1690$\pm$0.0081  \\
16 & 0.0026$\pm$0.0062 & 0.0209$\pm$0.0102 & 0.0443$\pm$0.0086 & 0.1280$\pm$0.0080  \\
17 & 0.0009$\pm$0.0040 & 0.0119$\pm$0.0085 & 0.0334$\pm$0.0082 & 0.0856$\pm$0.0073  \\
18 & 0.0001$\pm$0.0015 & 0.0068$\pm$0.0070 & 0.0261$\pm$0.0079 & 0.1640$\pm$0.0105  \\
19 & 0.0000$\pm$0.0000 & 0.0043$\pm$0.0060 & 0.0195$\pm$0.0075 & 0.1210$\pm$0.0101  \\
20 & 0.0000$\pm$0.0000 & 0.0010$\pm$0.0031 & 0.0129$\pm$0.0066 & 0.0755$\pm$0.0088  \\
21 & 0.0000$\pm$0.0000 & 0.0005$\pm$0.0024 & 0.0106$\pm$0.0064 & 0.0507$\pm$0.0079  \\
22 & 0.0000$\pm$0.0000 & 0.0000$\pm$0.0000 & 0.0054$\pm$0.0049 & 0.0287$\pm$0.0064  \\
23 & 0.0000$\pm$0.0000 & 0.0000$\pm$0.0000 & 0.0032$\pm$0.0041 & 0.0459$\pm$0.0086  \\
24 & 0.0000$\pm$0.0000 & 0.0000$\pm$0.0000 & 0.0027$\pm$0.0040 & 0.0271$\pm$0.0071  \\
25 & 0.0000$\pm$0.0000 & 0.0000$\pm$0.0000 & 0.0012$\pm$0.0028 & 0.0151$\pm$0.0057  \\
26 & 0.0000$\pm$0.0000 & 0.0000$\pm$0.0000 & 0.0002$\pm$0.0012 & 0.0159$\pm$0.0062  \\
27 & 0.0000$\pm$0.0000 & 0.0000$\pm$0.0000 & 0.0000$\pm$0.0000 & 0.0102$\pm$0.0052  \\
28 & 0.0000$\pm$0.0000 & 0.0000$\pm$0.0000 & 0.0000$\pm$0.0000 & 0.0125$\pm$0.0061  \\
29 & 0.0000$\pm$0.0000 & 0.0000$\pm$0.0000 & 0.0000$\pm$0.0000 & 0.0063$\pm$0.0046  \\
30 & 0.0000$\pm$0.0000 & 0.0000$\pm$0.0000 & 0.0000$\pm$0.0000 & 0.0057$\pm$0.0046  \\ \hline
\end{tabular}
\caption{Underdensity Probability Function $U(r)$ (using density threshold
  $\delta\rho/\rho=-0.8$) measured for volume-limited samples of the
  SGP region with absolute magnitude limits M$_{\rm lim}-5 {\rm
    log}h=-18, -19, -20$ and $-21$.  }
\label{tab:upfs}
\end{centering}
\end{table}


\end{document}